\RequirePackage{lineno}
\documentclass[aps,pre,notitlepage,longbibliography,groupedaddress]{revtex4-1}
\usepackage{graphicx}
\usepackage{epstopdf}
\usepackage{bm}
\usepackage{amssymb}
\usepackage{psfrag}
\usepackage{color}
\usepackage{amsbsy}      
\usepackage{amsmath,amsthm}
\usepackage{mathtools}
\usepackage{dsfont}
\usepackage{lineno}
\usepackage{comment}
\usepackage{hyperref}
\usepackage{tcolorbox}
\usepackage{booktabs}
\usepackage{url}
\usepackage{etoolbox}
\patchcmd{\section}
  {\centering}
  {\raggedright}
  {}
  {}
\patchcmd{\subsection}
  {\centering}
  {\raggedright}
  {}
  {}
\graphicspath{{./figures/}}
\usepackage{newfloat}
\DeclareFloatingEnvironment[name={Fig. D}]{suppfigure}
\usepackage{changes}

\def\dd{\mbox{d}}

\def\f{\frac}

\usepackage{comment}

\begin{document}

\title{Why case fatality ratios can be misleading: individual- and
  population-based mortality estimates and factors influencing them}



\author{Lucas B\"{o}ttcher$^{1}$}
\author{Mingtao Xia$^{2}$}
\author{Tom Chou$^{1,2,3}$}

\affiliation{$^{1}$Dept.~of Computational Medicine, UCLA, Los Angeles, CA 90095-1766}
\affiliation{$^{2}$Dept.~of Mathematics, UCLA, Los Angeles, CA 90095-1555}
\affiliation{$^{3}$Beijing Computational Science Research Center, Beijing, China}


\begin{abstract}
Different ways of calculating mortality ratios during epidemics
have yielded very different results, particularly during the current COVID-19
pandemic.  
%
%
%
We formulate both a survival probability model and an associated
infection duration-dependent SIR model to define individual- and
population-based estimates of dynamic mortality ratios. The key
parameters that affect the dynamics of the different mortality
estimates are the incubation period and the time individuals
were infected before confirmation of infection. We stress that none
of these ratios are accurately represented by the often misinterpreted
case fatality ratio (CFR), the number of deaths to date divided by the
total number of confirmed infected cases to date. Using data on the
recent SARS-CoV-2 outbreaks, we estimate and
compare the different dynamic mortality ratios and highlight their
differences. Informed by our modeling, we propose more systematic
methods to determine mortality ratios during epidemic outbreaks and
discuss sensitivity to confounding effects and uncertainties in the data.
%
\end{abstract}
\maketitle
\section*{Introduction}
\vspace{-2mm}

The mortality ratio is a key metric describing the severity of a viral
disease \cite{SEVERITY_LANCET}.  These metrics typically change in
time before converging to a constant value and can be defined in a
number of ways.  One commonly used metric is the (confirmed) case fatality ratio
(CFR or cCFR), the total number of deaths to date, $D(t)$, divided by
the total number of all confirmed cases to date
$N(t)$~\cite{H1N1_BMJ,xu2020pathological,wucharacteristics,SEVERITY_LANCET}.
Infection fatality ratios (IFR), the number of deaths to date divided
by the number of all infecteds, have also been used
\cite{REAL_TIME_JCM,CFR_IFR,CEBM_OXFORD} although the IFR requires an
estimate of the number of unconfirmed infecteds.

Both the CFR and IFR have been widely estimated from aggregated
population data from past outbreaks \cite{H1N1_BMJ} as well as from
those of the recent SARS-CoV-2 outbreaks
\cite{SEVERITY_LANCET,CEBM_OXFORD,REAL_TIME_JCM,JORDAN,CFR_CHINA,RUAN2020,SPYCHALSKI2020}. We
show examples of CFR curves in Fig.~\ref{fig:mortalitydata} and in the
\emph{Supplemental Information} (SI). As of March 31, 2020, the global
${\rm CFR}(t)=D(t)/N(t) = 42,158/858,892 \approx
4.9\%$~\cite{corona1}, while CFRs in individual regions vary
significantly. Clearly, this estimate would correspond to the actual
mortality ratio if (i) all infecteds were tested and (ii) \textit{all}
remaining unresolved individuals recover.  However, some infected
patients will die, increasing the estimated mortality ratio over
time. Despite the underestimation of this type of population-based
measurement, it is still commonly being used by various health
officials and is often inconsistently defined as deaths/(deaths +
recovereds) even though this difference has been clearly
distinguished~\cite{ghani2005methods}.

\begin{figure}[htb]
\centering
\includegraphics[width=0.7\textwidth]{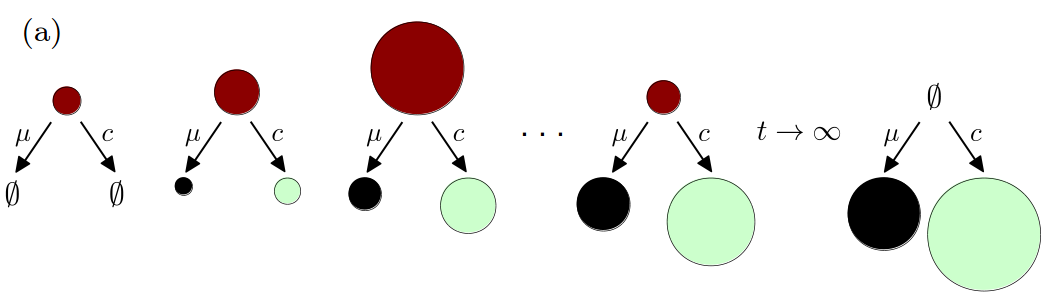}
\includegraphics{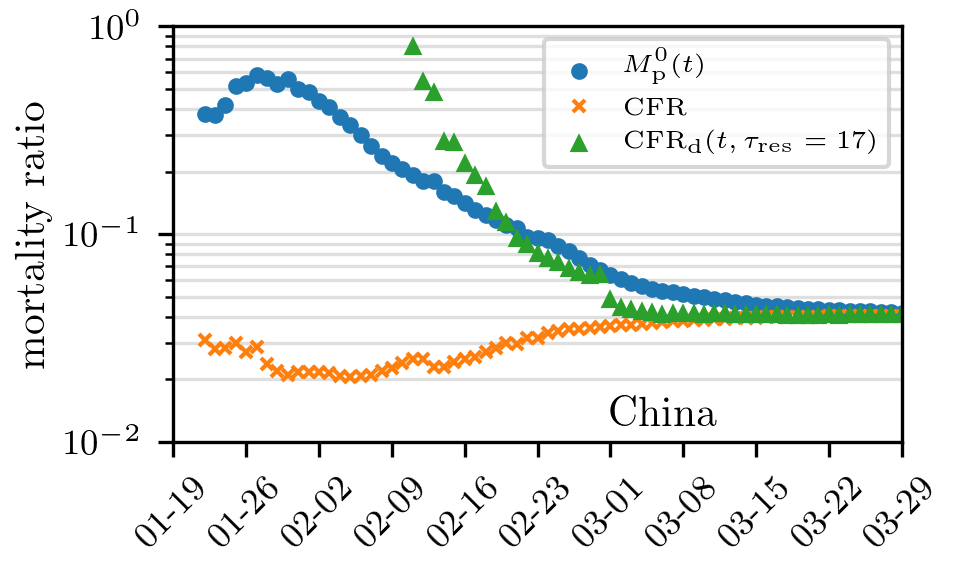}
\includegraphics{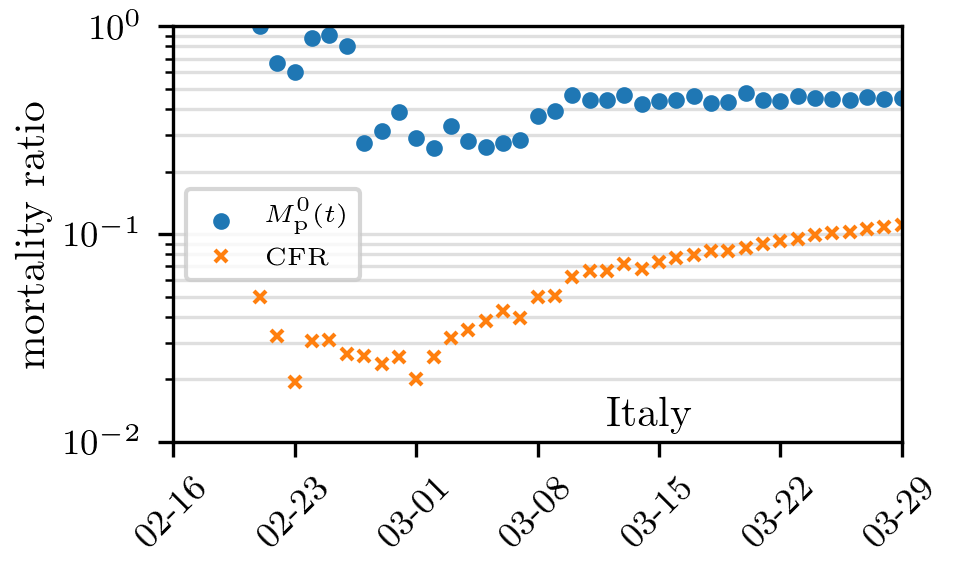}
\caption{\textbf{Mortality-ratio estimates}. (a) Evolution of the
  cumulative number of infected (red), death (black), and recovered
  (green) cases. The size of the circles indicates the number of cases
  in the respective compartments on a certain day. (b--c) Estimates of
  mortality ratios (see Eqs.~\eqref{eq:estimate2} and
  \eqref{eq:Mp0_Mp1}) of SARS-CoV-2 infections in China and Italy. The
  ``delayed'' mortality-ratio estimate $\mathrm{CFR}_\mathrm{d}$
  corresponds to the number of deaths to date divided by total number
  of cases at time $t-\tau_{\mathrm{res}}$. Many studies use
  $\mathrm{CFR}_\mathrm{d}$, although this metric underestimates the
  individual-based mortality (defined below).  Another
  population-based mortality ratio is $M_{\mathrm{p}}^0(t)$, the number
  of deaths divided by the sum of death and recovered cases, up to time
  $t$. The data are based on Ref.~\cite{dong2020interactive}.}
\label{fig:mortalitydata}
\end{figure}

During the severe 2003 acute respiratory syndrome (SARS) outbreak in
Hong Kong, the World Health Organization (WHO) also used the
aforementioned estimate to obtain a CFR of 4.5\% while the final
values, after resolution of infecteds, approached
17.0\%~\cite{yip2005comparison,yip2005chain}. For the ongoing
SARS-CoV-2 outbreaks, analyses by WHO and many others still use the
${\rm CFR}=D(t)/N(t)$ metric
\cite{SPYCHALSKI2020,CFR_CHINA,CEBM_OXFORD}, (see
Table~\ref{tab:M_estimates}), or time-shifted variants
\cite{CFR_IFR,SEVERITY_LANCET}.  Since true mortality rates are
critical for assessing the risks associated with epidemic outbreaks,
typical underestimations by ${\rm CFR}$s may lead to insufficient
countermeasures and a more severe
epidemic~\cite{bottcher2015disease,bottcher2016connectivity}.

An unambiguous definition of mortality ratio is the probability that a
single, newly infected individual will eventually die of the
disease. If there are sufficient individual-level or cohort data,
these probabilities can be further stratified according to patient
age, gender, health condition,
etc.~\cite{RIOU_2020,SEVERITY_LANCET}. This intrinsic mortality ratio,
or probability of death, should be an \textit{intrinsic} property of
the virus and the infected individual, depending on age, health,
access to health care, etc., and ought not \textit{directly} depend on
the population-level dynamics of infected and recovered individuals.
Thus, it can be framed as a survival probability of a single infected
individual.  Whether this individual infects others does not directly
affect his probability of eventually dying \footnote{Of course, at the
  population level, if there are many deaths, medical facilities are
  stressed which indirectly leads to an increase in death rates.}.

In the Results, we derive a model describing the probability
$M_{1}(t)$ that an infected individual dies or recovers before time
$t$. Importantly, these models incorporate the duration of infection
(including an incubation period) before a patient tests positive at
time $t=0$.  However, the CFR and other mortality measures are
typically reported based on population data. Do these population-based
measures, including CFR, provide reasonable measures of the
probability of death of an individual?  Also in the Results, we
describe how mortality ratios are defined within population-level
models, specifically, a disease duration-structured SIR model. We will
show that population-based estimates are typically not a meaningful
measure of mortality, but that under simplifying assumptions, the
mortality ratio $M_{\rm p}(t) = D(t)/(D(t)+R(t))$, where $R(t)$ is the
number of recovereds up to time $t$, is more closely related to the
true probability of death $M_{1}(t)$~\cite{ghani2005methods}.  The
simplest population-level mortality ratio is currently (as of March
31, 2020) $42,158/(42,158+178,100)\approx 19.1\%$~\cite{corona1},
significantly higher than the March 31, 2020 CFR$\approx 4.9\%$
estimate.

We use the same estimates for the rate parameters in our individual
and population models to compute the different mortality ratios. Note
that in general, both the individual mortality probability $M_{1}(t)$
and the population-based estimates $M_{\rm p}(t)$ depend on the time
of measurement $t$. By critically analyzing these estimates, the CFR,
and a ``delayed'' case fatality ratio ${\rm CFR}_{\rm d}$, we
illustrate and interpret the differences among these measures and
discuss how changes or uncertainty in the data affect them.  In
the Discussion, we summarize our results and identify a
correction factor to transform population-level mortality estimates
into individual mortality probabilities.
\begin{table}
\begin{tabular}{ll}
\hline
 \textbf{reference} & \textbf{CFR} \\\hline
 Xu \emph{et al.}~\cite{xu2020pathological,xu2020full} and Mahase~\cite{mahase2020coronavirus} & 2\% \\
 Wu \emph{et al.}~\cite{wucharacteristics} &  0.1-1\% (outside Wuhan) \\
  World Health Organization~\cite{whocovid,whocovid2} &  2-4\% \\
Porcheddu \textit{et al.}~\cite{porcheddu} & 2.3\% (Italy and China) \\
Peeri~\cite{peeri} \textit{et al.} & 2\% \\
\hline
\end{tabular}
\caption{Different CFR estimates of COVID-19.}
\label{tab:M_estimates}
\end{table}

\section*{Results}
\vspace{-2mm}

\subsection*{Intrinsic individual mortality rate}

\vspace{-4mm}
Consider an individual that, at the time of positive testing $(t=0)$,
had been infected for a duration $\tau_{1}$. A ``survival''
probability density can be defined such that $P(\tau, t\vert
\tau_{1})\dd \tau$ is the probability that the patient is still alive
and infected (not recovered) at time $t>0$ and has been infected for a
duration between $\tau$ and $\tau +\dd \tau$.  Since $\tau_{1}$ is
unknown, it must be estimated or averaged over some distribution.
%
%
%
%
The individual survival probability evolves according to 

\begin{equation}
{\partial P(\tau, t\vert \tau_{1})\over  \partial  t}
+ {\partial P(\tau, t\vert \tau_{1})\over  \partial  \tau}
 = -(\mu(\tau, t\vert \tau_{1})+c(\tau, t\vert \tau_{1}))P(\tau, t\vert \tau_{1}),
\label{eq:ind_survival}
\end{equation}
where the death and recovery rates, $\mu(\tau, t\vert \tau_{1})$ and
$c(\tau, t\vert \tau_{1})$, depend explicitly on the duration of
infection at time $t$ and implicitly on patient health and age
$a$ \footnote{The timescale of infections are much smaller than the
  aging process. In other words, the aging process across the disease
  time scale is negligible and the age-dependent transport terms
  $\partial P/\partial a$ can be neglected.}. They may also depend
explicitly on time $t$ to reflect changes in clinical policy or
available health care.  For example, enhanced medical care may
decrease the death rate $\mu$, giving the individual's intrinsic
physiological processes a chance to cure the patient. 


If we assume an initial condition of one individual having been
infected for time $\tau_{1}$ at the time of confirmation,
Eq.~\ref{eq:ind_survival} can be solved using the method of
characteristics (see the SI).
%
%
From the solution $P(\tau=t+\tau_{1}, t\vert \tau_{1})$ (see the
\textit{Methods}) one can derive the probabilities of death and
recovery by time $t$ as

\begin{equation}
P_{\rm d}(t\vert \tau_{1}) = \int_{0}^{t}\dd s\, \mu(\tau_{1}+s,s)
P(\tau_{1}+s, t\vert \tau_{1}),\quad 
P_{\rm r}(t\vert \tau_{1}) = \int_{0}^{t}\dd s\, c(\tau_{1}+s, s)
P(\tau_{1}+s, t\vert \tau_{1}).
\label{PDPC}
\end{equation}

%
The probability that an individual died before time $t$, 
conditioned on resolution (either death or recovery), is
then defined as

\begin{equation}
M_{1}(t\vert \tau_{1}) = {P_{\rm d}(t\vert \tau_{1}) \over P_{\rm d}(t\vert \tau_{1})
+P_{\rm r}(t\vert \tau_{1})}.
\label{eq:mortality_estimate1}
\end{equation}
%
%
Equations \eqref{PDPC} and \eqref{eq:mortality_estimate1} also
depend on all other relevant patient attributes such as age,
accessibility to health care, etc. In the long-time limit, when
resolution has occurred ($P_{\rm d}(\infty) +P_{\rm r}(\infty) = 1$),
the individual mortality ratio is simply $M_{1}(\infty) = P_{\rm
  d}(\infty)$. 
%
%
In order to capture the dependence of death and recovery rates on the
time an individual has been infected, we propose a constant recovery
rate $c$ and a piece-wise constant death rate $\mu(\tau \vert
\tau_{1})$ that is not explicitly a function of time $t$:

\begin{equation}
c(\tau,t\vert \tau_{1}) = c, \quad \mu(\tau \vert\tau_{1}) =
\left\{ \begin{array}{cl} 0 & \tau \leq \tau_{\rm inc}
  \\ \mu_{1} & \tau > \tau_{\rm inc} \end{array} \right.,
\label{eq:recovery_mortality_rates}
\end{equation}
where $\tau_{\rm inc}$ is the incubation time during which the patient
is asymptomatic, has zero death rate, but can recover by clearing the
virus. In other words, some patients fully recover without ever
developing serious symptoms.

\begin{figure}
\includegraphics{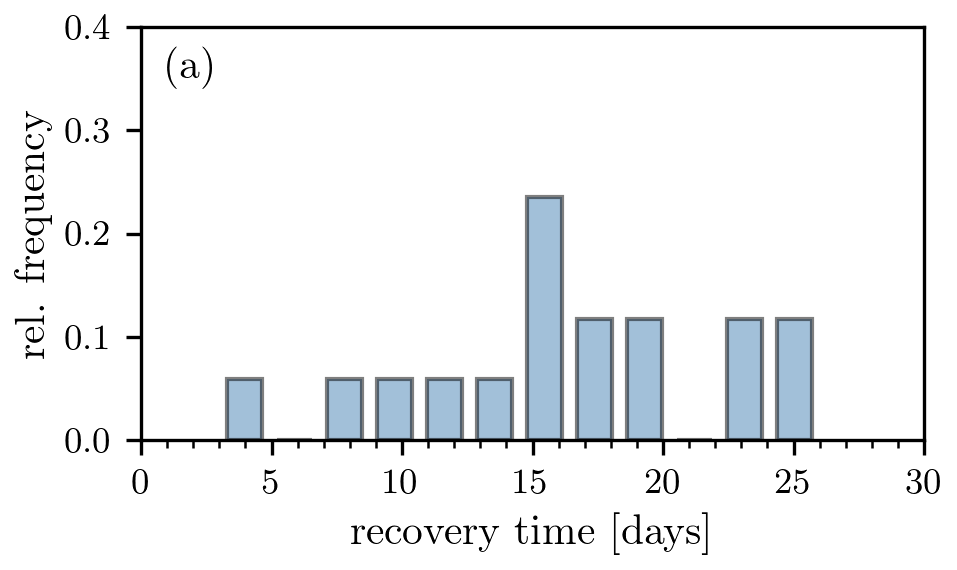}
\includegraphics{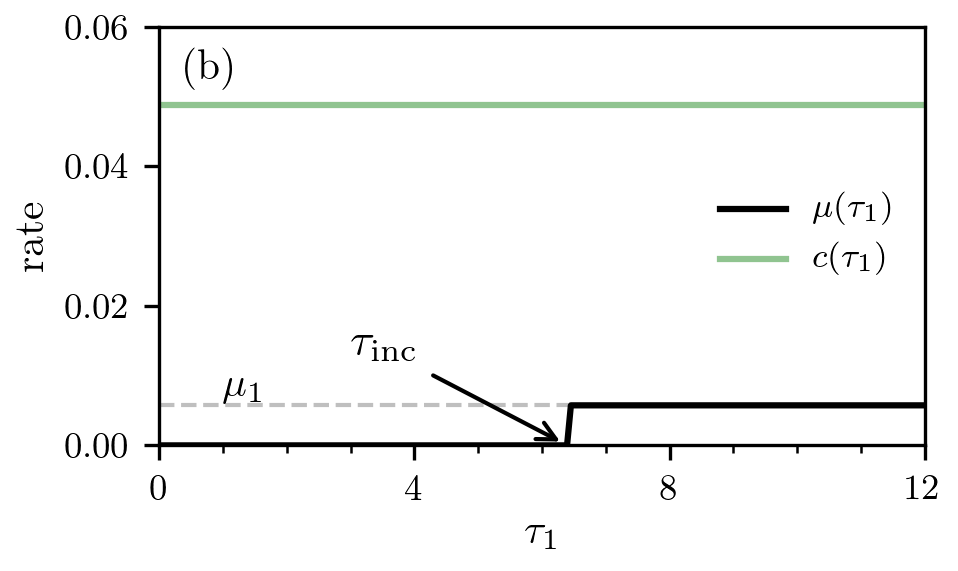}
\includegraphics{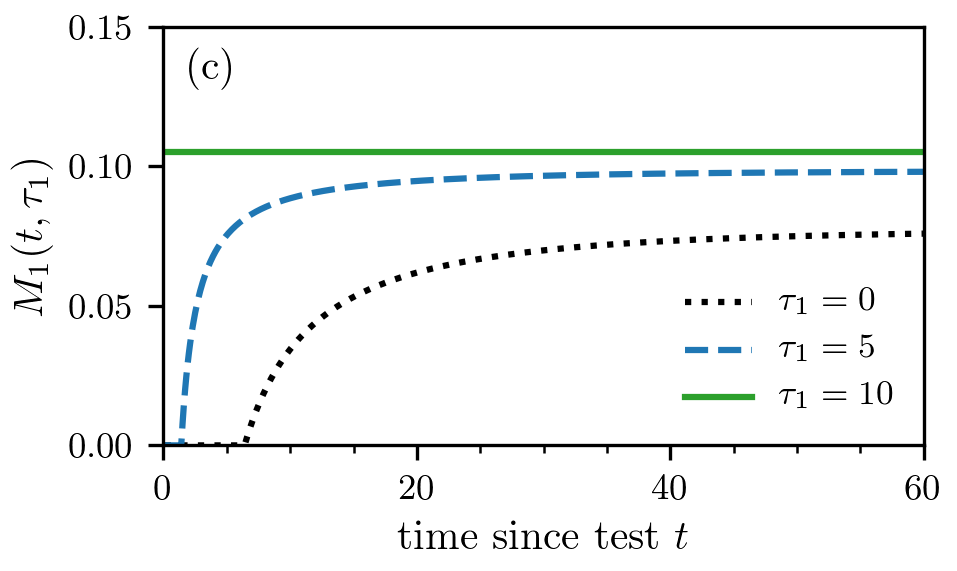}
\includegraphics{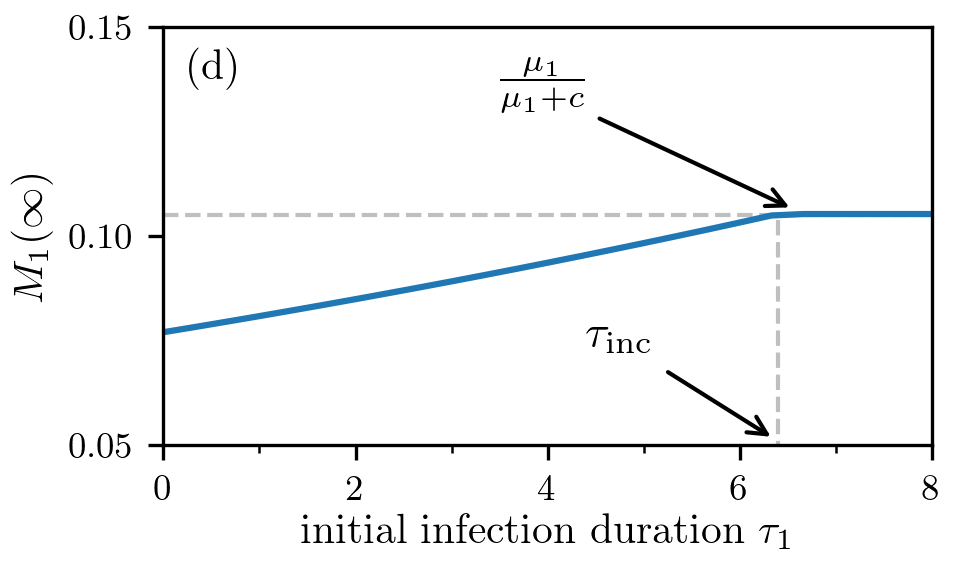}
\caption{\textbf{Individual mortality ratio}. (a) Recovery time after
  first symptoms occurred based on individual data of 17
  patients~\cite{kraemer2020epidemiological}. (b) Death- and recovery
  rates as defined in Eq.~\eqref{eq:recovery_mortality_rates}. The
  death rate $\mu(\tau_1)$ approaches $\mu_1$ for $\tau_1 >
  \tau_\mathrm{inc}$, where $\tau_\mathrm{inc}$ is the incubation
  period and $\tau_1$ is the time the patient has been infected before
  first being tested positive. (c) The individual mortality ratio
  $M_1(t\vert \tau_{1})$ for $\tau_{\rm inc}=6.4$ days at different
  values of $\tau_1$. Note that the individual death probability
  $P_{\mathrm{d}}(t\vert\tau_{1})$ and $M_1(t\vert \tau_{1})$ are
  nonzero only after $t>\tau_{\rm inc}-\tau_{1}$. (d) The asymptotic
  individual mortality ratio $M_1(\infty)$ (see
  Eq.~\eqref{eq:mortality_estimate1}) as a function of $\tau_1$.}
\label{fig:individual_estimate}
\end{figure}
For coronavirus infections, the incubation period appears to be highly
variable with a mean of $\tau_{\rm inc}\approx
6.4~\text{days}$~\cite{lai2020severe}.  We can estimate $\mu_1$ and
$c$ using individual patient data where 19 patients (outside Hubei)
had been tracked from the date on which their first symptoms occurred
until the disease resolved~\cite{kraemer2020epidemiological}.

Two out of 19 patients died, on average, 20.5 days after first
symptoms occurred and the mean recovery time of the remaining 17
patients is 16.8 days. We show the recovery-time distribution in
Fig.~\ref{fig:individual_estimate}(a). Since we know that the
mortality ratio in this dataset is $2/19$, we can determine the
dependence between $\mu_1$ and $c$ according to $\mu_1/(\mu_1+c)
\approx 2/19$ (or $c/\mu_1 \approx 8.5$).  The constant recovery and
after-incubation period death rates~\cite{keeling2011modeling} are
estimated to be
\begin{equation}
c = \frac{1}{20.5}/\text{day}\approx 0.049/\text{day} 
\quad \text{and}\quad \mu_1 = c / 8.5 \approx 0.006/\text{day}.
\label{eq:rates}
\end{equation}

Using these numbers, the recovery and death rate functions
$c(\tau,t\vert \tau_{1})$ and $\mu(\tau \vert\tau_{1})$ are plotted as
functions of $\tau$ in Fig.~\ref{fig:individual_estimate}(b). We show
the evolution of $M_{1}(t\vert \tau_{1})$ at different values of
$\tau_1$ in Fig.~\ref{fig:individual_estimate}(c). The corresponding
long-time limit $M_{1}(\infty)$ is readily apparent in
Fig.~\ref{fig:individual_estimate}(d): for $\tau_1 \geq
\tau_\mathrm{inc}$, $M_{1}(\infty)=\mu_1/(\mu_1+c)\approx 0.105$,
while $M_{1}(\infty) < \mu_1/(\mu_1+c)$ when $\tau_1 <
\tau_\mathrm{inc}$. The smaller expected mortality associated with
early identification of infection arises from the remaining incubation
time during which the patient has a chance to recover without
possibility of death. When conditioned on testing positive at or after
the incubation period, the patient immediately suffers a positive
death rate, increasing his $M_{1}(\infty)$.

Finally, in order to infer $M_{1}$ (and also indirectly $\mu$ and $c$)
during an outbreak, a number of statistical issues must be considered.
First, if the outbreak is ongoing, there may not be sufficient
long-time cohort data. Second, $\tau_{1}$ is unknown. Since testing
typically occurs at the onset of symptoms, most positive patients will
have been infected a few days earlier. The uncertainty in $\tau_{1}$
can be represented by a probability density $\rho(\tau_{1})$ for the
individual. The expected mortality can then be constructed as an
average over $\rho(\tau_{1})$:

\begin{equation}
\bar{M}_{1}(t) = {\bar{P}_{\rm d}(t) \over \bar{P}_{\rm d}(t)+\bar{P}_{\rm r}(t)},
\label{eq:M1bar}
\end{equation}
where $\bar{P}_{\rm d}(t)$ and $\bar{P}_{\rm r}(t)$ are the
$\tau_{1}$-averaged probabilities death and cure probabilities.

%
%

Some properties of the distribution $\rho(\tau_1)$ can be inferred
from the behavior of patients. Before symptoms arise, only very few
patients will know they have been infected, seek medical care, and get
their case confirmed (\textit{i.e.}, $\rho(\tau_1) \approx 0 $ for
$\tau_1\approx 0$). The majority of patients will contact
hospitals/doctors when they have been infected for a duration of
$\tau_{\rm inc}$. The distribution $\rho(\tau_1)$ thus reaches its
maximum near or shortly after $\tau_{\rm inc}$. Since patients are
most likely to test positive after experiencing symptoms, we choose a
gamma distribution

\begin{equation}
\rho(\tau_1; n, \gamma)= \frac{\gamma^n}{\Gamma(n)} \tau_1^{n-1} e^{-\gamma \tau_1}
\label{eq:gamma}
\end{equation}
with shape parameter $n=8$ and rate parameter $\gamma=1.25/\rm{day}$
so that the mean $n/\gamma $ is equal to $\tau_{\rm inc}=6.4$. 
%
%

Upon using the rates in Eqs.~\eqref{eq:recovery_mortality_rates} and
averaging over $\rho(\tau_{1})$, we derived expressions for
$\bar{P}(t), \bar{P}_{\rm d}(t)$, and $\bar{P}_{\rm r}(t)$ which are
explicitly given in the SI. Using the values in Eq.~\eqref{eq:rates}
we find an expected individual mortality ratio $\bar{M}_{1}(t)$ (which
are subsequently plotted in Fig.~\ref{fig:sir}) and its asymptotic
value $\bar{M}_{1}(\infty) = \bar{P}_{\rm d}(\infty)=0.101$. Of
course, it is also possible to account for more complex time-dependent
forms of $c$ and $\mu_1$~\cite{bottcher2020unifying}, but we will
primarily use Eqs.~\eqref{eq:recovery_mortality_rates} in our
subsequent analyses.

In the next section, we define population-based estimates for
mortality ratios, $M_{\rm p}(t)$, and explore how they can be computed
using SIR-type models. By comparing $M_{1}(t)$ to $M_{\rm p}(t)$, we
gain insight into whether population-based metrics are good proxies
for individual mortality ratios. We will outline the mathematical
differences and additional errors that confound population-level
estimates.

%
%

\subsection*{Infection duration-dependent SIR model}
\label{SIR1}
\vspace{-4mm}

While individual mortalities can be estimated by tracking many
individuals from infection to recovery or death, oftentimes, the
available data are not resolved at the individual level and only total
populations are given. Typically, one has the total number of cases
accumulated up to time $t$, $N(t)$, the number of deaths to date
$D(t)$, and the number of cured/recovered patients to date $R(t)$ (see
Fig.~\ref{fig:mortalitydata}). The CFR is simply $D(t)/N(t)$.  Note
that $N(t)$ includes unresolved cases and that $N(t) \geq R(t) +
D(t)$. Resolution (death or recovery) of all patients, $N(\infty) =
R(\infty) +D(\infty)$, occurs only well after the epidemic passes.

A variant of the CFR commonly used in the
literature~\cite{xu2020pathological,wucharacteristics} is the delayed
CFR
\begin{equation}
{\rm CFR}_{\rm d}(t,\tau_{\rm res}) = \frac{D(t)}{N(t-\tau_{\rm res})},
\label{eq:estimate2}
\end{equation}
where $\tau_{\rm res}$ is a corresponding time lag that accounts for
the duration from the day when first symptoms occurred to the day of
cure/death. Many estimates of the COVID-19 mortality ratio assume that
$\tau_{\rm res}=0$~\cite{xu2020pathological,wucharacteristics} and
thus underestimate the number of death cases $D(t)$ that result from a
certain number of infected individuals. Similar underestimations using
${\rm CFR}_{\rm d}$ have been reported in previous epidemic outbreaks
of SARS~\cite{ghani2005methods,yip2005comparison} and
Ebola~\cite{atkins2015under}.

Alternatively, a simple and interpretable population-level mortality
ratio is $M_{\rm p}(t) = D(t)/(R(t) + D(t))$, the death ratio of all
\textit{resolved} cases. To provide a concrete model for $D(t)$ and
$R(t)$, and hence $M_{\rm p}(t)$, we will use a variant of the
standard infection duration-dependent susceptible-infected-recovered
(SIR)-type model described by~\cite{NATURE_PDE}

\begin{align}
{\dd S(t)\over \dd t}  & =  
-S(t)\int_{0}^{\infty}\!\dd\tau'\, \beta(\tau',t)I(\tau',t),\nonumber \\
{\partial I(\tau, t)\over \partial t}+{\partial I(\tau,t)\over
  \partial \tau} & =
-(\mu(\tau,t)+c(\tau,t))I(\tau,t),
%
\label{SIR1_EQNS}
\end{align}
and $\dd R(t)/\dd t = \int_{0}^{\infty}\!\dd \tau c(\tau,t)I(\tau,t)$,
where $S(t)$ is the number of susceptibles, $I(\tau,t)$ is density of
individuals at time $t$ who have been infected for time $\tau$, and
$R(t)$ is the number of recovered individuals.  The rate at which an
individual infected for time $\tau$ at time $t$ transmits the
infection to a susceptible is denoted by $\beta(\tau, t)S(t)$.  For
simplicity, we assume only community spread and neglect immigration of
infected, which can be straightforwardly incorporated
\cite{NATURE_PDE}.

Note that the equation for $I(\tau,t)$ is identical to the equation for
the survival probability described by Eq.~\eqref{eq:ind_survival}.  It
is also equivalent to McKendrick age-structured models
\cite{MCKENDRICK,GREENMANJSP}
\footnote{In both the individual model (Eq.~\eqref{eq:ind_survival})
  and population models (Eq.~\eqref{SIR1_EQNS}), the death and cure
  rates are insensitive to changes in age $a$ over the time scale of
  epidemic if it occurs over, say, only 1-2 years. In this limit, we
  consider only infection-duration dependence in the population
  dynamics.}. Infection of susceptibles is described by the boundary
condition

\begin{equation}
I(\tau = 0, t) = S(t)\int_{0}^{\infty}\!\!\dd \tau'\,
\beta(\tau', t)I(\tau', t),
\label{eq:bc_SIR}
\end{equation}
which is similar to that used in age-structured models to represent
birth~\cite{MCKENDRICK}. Finally, we use an initial condition
consistent with the infection duration density given by
Eq.~\eqref{eq:gamma}: $I(\tau,0) = \rho(\tau;n=8,\gamma=1.25)$. Note
that Eq.~\eqref{eq:bc_SIR} assumes that all newly infected individuals
are immediately identified; \textit{i.e.}, these newly infected
individuals start with $\tau_{1}=0$. After solving for the infected
population density, the total number of deaths and recoveries to date
can be found via

\begin{equation}
D_0(t) = \int_{0}^{t}\!\!\dd t' \int_{0}^{\infty}\!\!\dd \tau\,  
\mu(\tau, t')I(\tau,t'),\quad  R_0(t) =
\int_{0}^{t}\!\!\dd t'\int_{0}^{\infty}\!\!\dd \tau\,
c(\tau, t')I(\tau,t').
\label{eq:Nmu0_Nc0}
\end{equation}
The corresponding total number of cases $N(t)$ in Eq.~\eqref{eq:estimate2} is 
\begin{equation}
N_0(t)=R_0(t)+D_0(t) + \int_{0}^{\infty}\!\!\dd \tau\,
I(\tau, t).
\label{eq:NT0}
\end{equation}
In the definitions of $D_0(t)$, $R_0(t)$, and $N_0(t)$, we account for
all possible death and recovery cases to date (see SI) and that newly
infected individuals are immediately identified. We use these case
numbers as approximations of the reported case numbers to study the
evolution of mortality-ratio estimates. Mortality ratios based on
these numbers underestimate the actual individual mortality $M_1$ (see
the previous ``Intrinsic individual mortality rate'' subsection) since
they involve individuals that have been infected for different
durations $\tau$, particularly recently infected individuals who have
not yet died.

An alternative way to compute populations is to exclude the newly
infecteds and consider only the initial cohort. The corresponding
populations in this case are defined as
\begin{equation}
D_1(t) = \int_{0}^{t}\!\!\dd t'\int_{t'}^{\infty}\!\!\dd \tau\,
\mu(\tau, t')I(\tau,t'),\quad  R_1(t) =
\int_{0}^{t}\!\!\dd t' \int_{t'}^{\infty}\!\!\dd \tau\,
c(\tau, t')I(\tau,t').
\label{eq:Nmu1_Nc1}
\end{equation}
%
%
%
Since $D_1(t)$ and $R_1(t)$ do not include infecteds with $\tau < t$,
they exclude the effect of newly infected individuals, but may yield
more accurate mortality-ratios as they are based on an initial cohort
of individuals in the distant past. The infections that occur after
$t=0$ contribute only to $I(\tau < t, t)$; thus, $D_1(t)$ and $R_1(t)$
do not depend on the transmission rate $\beta$, possible immigration
of infecteds, or the number of susceptibles $S(t)$.  Note that all the
populations derived above implicitly average over $\rho(\tau_1; n,
\gamma)$ for the first cohort of identified infecteds (but not
subsequent infecteds). Moreover, the population density $I(\tau\geq t,
t)$ follows the same equation as $\bar{P}(t\vert \tau_{1})$ provided
the same $\rho(\tau_1; n, \gamma)$ is used in their respective
calculations.

The two different ways of partitioning populations
(Eqs.~\eqref{eq:Nmu0_Nc0} and \eqref{eq:Nmu1_Nc1}) lead to two
different population-level mortality ratios
\begin{equation}
M_\mathrm{p}^0(t) = {D_0(t)\over  D_0(t)+ R_{0}(t)} 
\quad \mathrm{and} \quad M_\mathrm{p}^1(t) = {D_1(t)\over D_1(t)+R_{1}(t)}.
\label{eq:Mp0_Mp1}
\end{equation}
Since the populations $D_{0}(t)$ and $R_{0}(t)$, and hence
$M_\mathrm{p}^{0}(t)$, depend on disease transmission through
$\beta(\tau,t)$ and $S(t)$, we expect $M_{\rm p}^{0}(t)$ to carry a
different interpretation from $M_{1}(t)$ and $M_{\rm p}^1(t)$.

In the special case in which $\mu$ and $c$ are constants, the
time-integrated populations $\int_{0}^{t}\!\dd
t'\,\int_{0}^{\infty}\!\dd \tau\, I(\tau,t')$ and $\int_{0}^{t}\!\dd
t'\,\int_{t'}^{\infty}\!\dd \tau\,I(\tau,t')$ factor out of $M_{\rm
  p}^0(t)$ and $M_{\rm p}^1(t)$, rendering them time-independent and

\begin{equation}
M_{\rm p}^{0,1} = {\mu_1 \over \mu_1 + c} = M_{1}.
\end{equation}
Thus, only in the special time-homogeneous case do both
population-based mortality ratios become \textit{independent} of the
population (and transmission $\beta$) and coincide with the individual
death probability.

\begin{figure}[htb]
\centering
\includegraphics{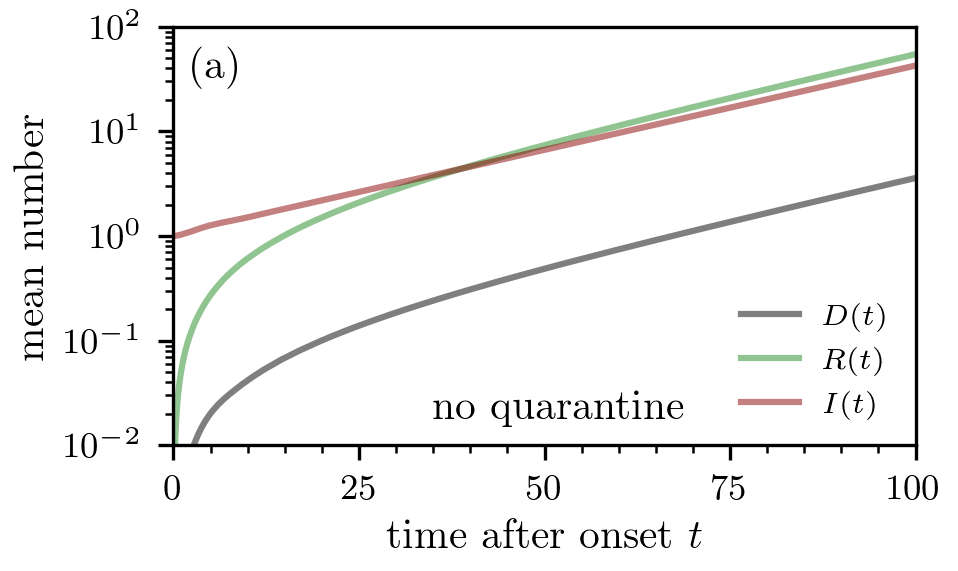}
\includegraphics{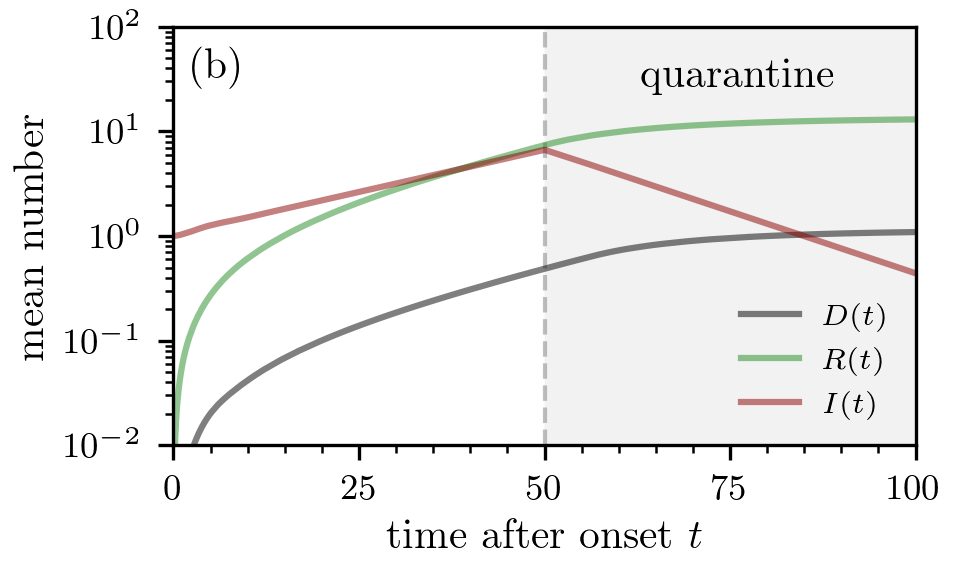}
\includegraphics{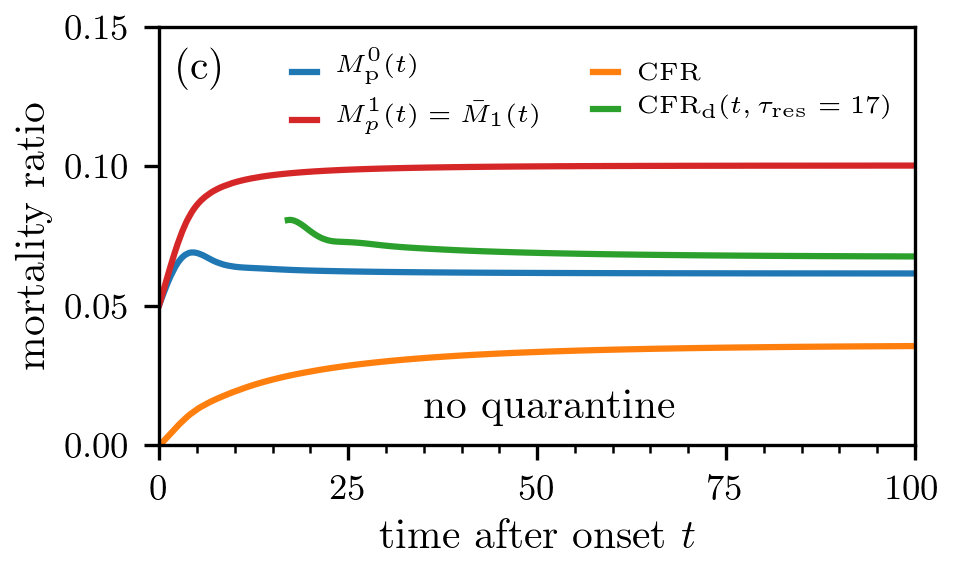}
\includegraphics{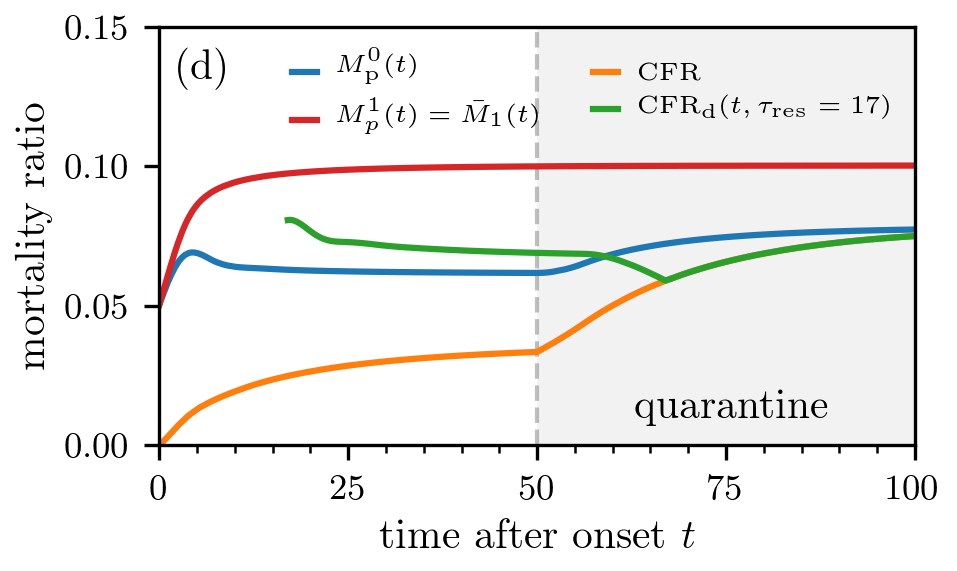}
\caption{\textbf{Population-level mortality-ratio estimates.}
  Outbreak evolution and mortality ratios without containment measures
  (a,c) and with quarantine (b,d). The curves are based on numerical
  solutions of Eqs.~\eqref{SIR1_EQNS} using the initial condition
  $I(\tau, 0) = \rho(\tau;8,1.25)$ (see Eq.~\eqref{eq:gamma}). The
  death and recovery rates are defined in
  Eqs.~\eqref{eq:recovery_mortality_rates} and \eqref{eq:rates}.  We
  use a constant infection rate $\beta_{1}S(0)=0.158$/day, which we
  estimated from the basic reproduction number of
  SARS-CoV-2~\cite{lai2020severe}. To model quarantine effects, we set
  $\beta_{1}=0$ for $t>50$. We show the mortality-ratio estimates
  $M_\mathrm{p}^0(t)$ and $M_\mathrm{p}^1(t)$ (see
  Eq.~\eqref{eq:Mp0_Mp1}) and
  $\mathrm{CFR}_{\mathrm{d}}(t,\tau_{\mathrm{res}})$ (see
  Eqs.~\eqref{eq:estimate2}, \eqref{eq:Nmu0_Nc0}, \eqref{eq:NT0}, and
  \eqref{eq:Mp0_Mp1}). }
\label{fig:sir}
\end{figure}

To illustrate the differences between $M_{1}(t), M_{\rm p}^{0,1}(t)$,
and ${\rm CFR}_{\rm d}(t,\tau_{\rm res})$ in more general cases, we
use the simple death and cure rate functions given by
Eqs.~\eqref{eq:recovery_mortality_rates} in solving
Eqs.~\eqref{eq:ind_survival} and \eqref{SIR1_EQNS}. For
$\beta(\tau,t)$ in Eq.~\eqref{eq:bc_SIR}, we account for incubation
effects by neglecting transmission during the asymptomatic incubation
period ($\tau \leq \tau_{\rm inc}$) and assume
\begin{equation}
\beta(\tau, t) = \left\{ \begin{array}{cl} 0 & \tau \leq \tau_{\rm inc} \\
\beta_{1} & \tau > \tau_{\rm inc}. \end{array} \right. 
\end{equation}
We use the estimated basic reproductive number ${\cal R}_{0} =
\beta_{1}S(0)/(\mu_1+c) \approx 2.91$~\cite{lai2020severe} to fix
$\beta_{1}S(0) = (\mu_1+c){\cal R}_{0} \approx 0.158/{\rm day}$. We
also first assume that the susceptible population does not change
appreciably before quarantine and set $S(t) = S(0)$. Thus, we only
need to solve for $I(\tau,t)$ in Eqs.~\eqref{SIR1_EQNS} and
\eqref{eq:bc_SIR}. We solve Eqs.~\eqref{SIR1_EQNS} and
\eqref{eq:bc_SIR} numerically (see the \emph{Methods} section for
further details) and use these numerical solutions to compute
$D_{0,1}(t)$, $R_{0,1}(t)$, and $N_{0,1}(t)$ (see
Fig.~\ref{fig:sir}(a) and (b)), which are then used in
Eqs.~\eqref{eq:Mp0_Mp1} and ${\rm CFR}_{\rm d}(t-\tau_{\rm res})$. To
determine a realistic value of the time lag $\tau_{\rm res}$, we use
data on death/recovery periods of 36 tracked
patients~\cite{kraemer2020epidemiological} and find that patients
recover/die, on average, $\tau_{\rm res}=16.5~\text{days}$ after first
symptoms occurred.

We show in Figs.~\ref{fig:sir}(c) and (d) that $M_\mathrm{p}^{1}(t)$
approaches the individual mortality ratio $\bar{M}_1(\infty)\approx
0.1$ given in the ``Intrinsic individual mortality rate'' subsection
above. This occurs because the model for $P(\tau, t)$ and $I(\tau,t)$
are equivalent and we assumed the same initial distribution
$\rho(\tau;8,1.25)$ for both quantities.  However, the
population-level mortality ratios ${\rm CFR}_\mathrm{d}(t,\tau_{\rm
  res})$ and $M_\mathrm{p}^{0}(t)$ also take into account recently
infected individuals who may recover before symptoms. This difference
yields different mortality ratios because newly infecteds are
implicitly assumed to be detected immediately and all have
$\tau_{1}=0$.  Thus, the underlying infection-time distribution is not
the same as that used to compute $\bar{M}_\mathrm{p}^{1}(t)$ (see the
SI for further details).  The mortality ratios ${\rm
  CFR}_\mathrm{d}(t,\tau_{\rm res})$ and $M_\mathrm{p}^{0}(t)$ should
not be used to quantify the individual mortality probability of
individuals who tested positive.  Moreover, due to evolution of the
disease, $D(t)$, $R(t)$, and $N(t)$ do not change with the same rates
during an outbreak, the population-level mortality measures ${\rm
  CFR}_\mathrm{d}(t,\tau_{\rm res})$ and $M_\mathrm{p}^{0}(t)$ reach
their final steady state values only after sufficiently long times
(see Fig.~\ref{fig:sir}(c) and (d)).


The evolution of the mortality ratios in Fig.~\ref{fig:sir}
qualitatively resembles the behavior of the mortality-ratio estimates
in Fig.~\ref{fig:mortalitydata}. As shown in
Fig.~\ref{fig:mortalitydata}, the population-based estimates for
coronavirus varies, decreasing in time for China but fluctuating for
Italy. These changes could result from changing practices in data
collecting, or from explicitly time-inhomogeneous parameters
$\mu(\tau, t)$, $c(\tau, t)$, and/or $\beta(\tau,t)$.

Although population-level quarantining does not directly affect the
individual mortality $M_{1}(t\vert \tau_{1})$ or $\bar{M}_{1}(t)$, it
can be easily incorporated into the SIR-type population dynamics
equations through changes in $\beta(\tau, t)S(t)$.  For example, we
have set $S(t>t_{\mathrm{q}}) = 0$ to represent implementation of a
quarantine after $t_{\mathrm{q}}=50$ days of the outbreak. After
$t_{\mathrm{q}}=50$ days, no new infections occur and the estimates
${\rm CFR}_\mathrm{d}(t,\tau_{\rm res})$ and $M_\mathrm{p}^{0}(t)$
start converging immediately towards their steady-state values (see
Fig.~\ref{fig:sir}(d)). Since the number of deaths decreases after the
implementation of quarantine measures, the delayed ${\rm
  CFR}_\mathrm{d}(t,\tau_{\rm res}=17)$ is first decreasing until
$t=t_{\mathrm{q}}+\tau_{\rm res}=67$. For $t > 67$, the ${\rm
  CFR}_\mathrm{d}(t,\tau_{\rm res}=17)$ measures no new cases and is
thus equal to the CFR.
\section*{Discussion and Summary}
\label{sec:discussion}
\vspace{-2mm}

%


%
\begin{figure}
\centering 
\includegraphics[width=3.2in]{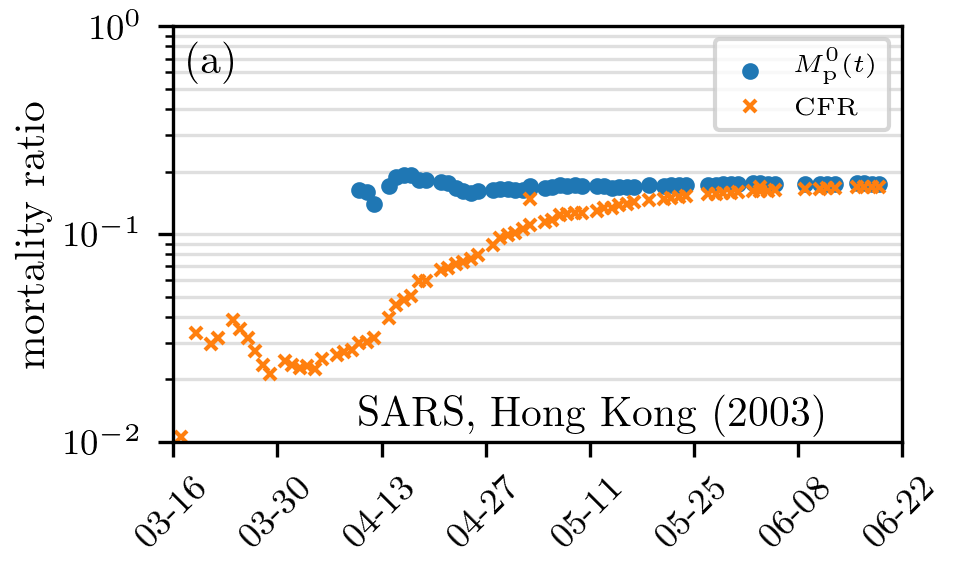}
\includegraphics[width=3.2in]{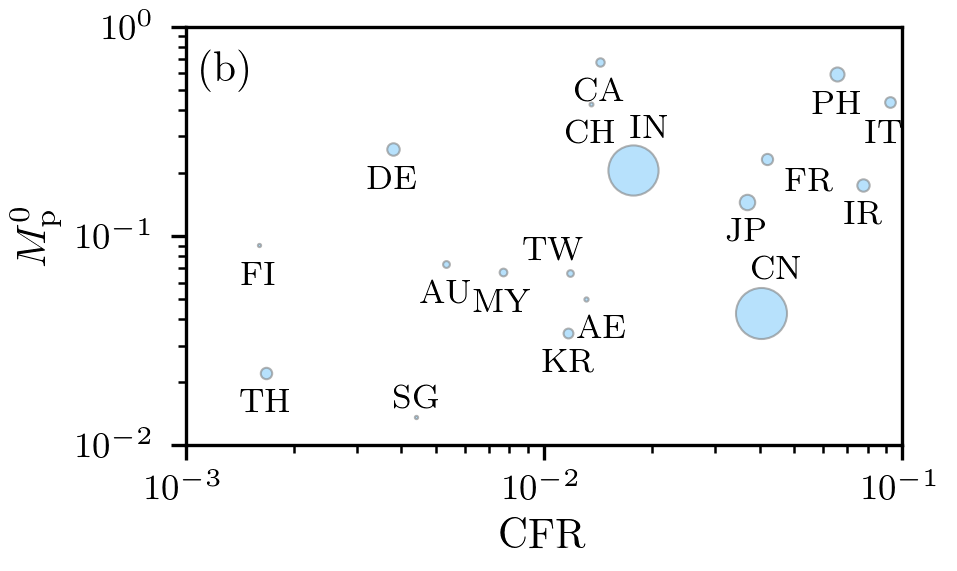}
\caption{\textbf{SARS mortality and region-dependence of COVID-19
    mortality-ratio estimates.} (a) Estimates of mortality ratios (see
  Eqs.~\eqref{eq:estimate2} and \eqref{eq:Mp0_Mp1}) of SARS infections
  in Hong Kong (2003)~\cite{whoSARS}.  (b) Mortality-ratio estimates
  of COVID-19 in different regions (see Eqs.~\eqref{eq:estimate2} and
  \eqref{eq:Mp0_Mp1} ($\tau_{\rm res}=0$)). We used data on the
  cumulative number of cases, recoveries, and deaths in
  Ref.~\cite{dong2020interactive} as of March 24, 2020. The marker
  sizes indicate the population of the corresponding countries. The
  metrics $M_{\rm p}^{0}(t)$ and and CFR are largely uncorrelated with
  correlation coefficient 0.33.}
\label{fig:pop_data}
\end{figure}

During an epidemic, it is important to assess the severity of the
disease by estimating its mortality and other disease
characteristics. Assuming accurate data, the often-used CFR typically
underestimates the true, final death ratio. For example, during the
SARS outbreaks in Hong Kong, the WHO first estimated the fatality rate
to 2.5\% (March 30, 2003) whereas the final estimates reached values
of about 17.0\% (June 30, 2003) (see
Fig.~\ref{fig:pop_data}(a))~\cite{yip2005chain}. Standard metrics like
the CFR are seen to be easily confounded by and sensitive to
uncertainty in intrinsic disease parameters such as the incubation
period and the time $\tau_{1}$ a patient had been infected before
clinical confirmation of infection. For the recent COVID-19 outbreaks,
CFR-based measures may still provide reasonable estimates of the
actual mortality across different age classes due to a counter-acting
error in the numbers of unreported mild-symptom cases.

Here, we stress that more mechanistically meaningful and interpretable
metrics can be defined and, as easily, estimated from population data as
CFRs. Our proposed mortality ratios for viral epidemics are defined in
terms of (i) individual survival probabilities and (ii) population
ratios using numbers of deaths and recovered individuals. Both of
these measures are based on the within-host evolution of the disease,
and in the case of $M_{\rm p}^{0,1}(t)$, the population-level
transmission dynamics.  
%

Among the metrics we describe, $M_{\rm p}^{1}(t)$ is structurally
closest to the individual mortality $\bar{M}_{1}(t)$ in that both are
independent of disease transmission since new infections are not
considered. Both of these mortality ratios converge after an
incubation time $\tau_{\rm inc}$ to a value smaller than or equal to
$\mu_1/(\mu_1+c)$. 


The most accurate estimates of $M_{1}$ can be obtained if we keep
track of the fate of cohorts that were infected within a small time
window in the past. By following only these individuals, one can track
how many of them died as a function of time. As more cases arise, one
should stratify them according to their estimated times since
infection to gather improving statistics for $M_{1}(\infty)$.  With
the further spread of SARS-CoV-2 in different countries, data on more
individual cases of death and recovery can be more easily stratified
according to other central factors in COVID-19 mortality: age, sex,
health condition.

Besides accurate cohort data, for which at present there are few for
coronavirus, cumulative population data has been used to estimate the
mortality ratio. The population-level metrics $M_\mathrm{p}^{0}(t)$
and ${\rm CFR}(t)$ implicitly depend on new infections and the
transmission rate $\beta$. Despite this confounding factor,
$M_\mathrm{p}^{0}(t)$ and ${\rm CFR}_{\rm d}(t,\tau_{\rm res})$
approach $e^{-c \tau_{\rm inc}} \mu_1/(\mu_1+c)$ as $t\rightarrow
\infty$, where $e^{-c \tau_{\rm inc}}$ is the probability that no
recovery occurred during the incubation time $\tau_{\rm inc}$. Based
on these results, we can establish the following connection between
the different mortality ratios for initial infection times with
distribution $\rho(\tau_{1};n,\gamma)$ and mean $\bar{\tau}=n/\gamma$:
\begin{equation}
{\rm CFR}_{\rm
    d}(\infty)=M_{\rm p}^0(\infty) \approx e^{- c \bar{\tau}} M_{\rm
    p}^1(\infty)=e^{- c \bar{\tau}} \bar{M}_1(\infty)\,.
\label{eq:connection}
\end{equation}
%
%
According to Eq.~\eqref{eq:connection}, population-level mortality
estimates (\textit{e.g.}, CFR and $M_{\rm p}^0$) can be transformed,
at least approximately, into individual mortality probabilities using
the correction factor $e^{- c \bar{\tau}}$ with $\bar{\tau}\approx
\tau_{\rm inc}$.

Besides the mathematical differences between $M_{1}(t)$ and $M_{\rm
  p}^{0}(t)$, CFR, estimating $M_{\rm p}^{0}(t)$ and ${\rm CFR}(t)$
from aggregate populations implicitly incorporate a number of
confounding factors that contribute to their variability. In
Fig.~\ref{fig:pop_data}(b), we plot the population-level mortality-ratio
estimates $M_\mathrm{p}^{0}$ against the CFR for different regions and
observe large variations and very little correlation between
countries~\cite{battegay20202019}. As of March 31, 2020, the value of
$M_{\rm p}^0$ in Italy is almost 45\% and can increase further if the
current conditions (e.g., treatment methods, age group proportion of
infecteds, etc.) do not change. Differences between the mortality
ratios in China and Italy (see Figs.~\ref{fig:mortalitydata}(b) and
(c)) might be a result of varying medical treatment strategies,
different practices in data collecting (e.g., post-mortem testing),
differences in the age demographics between the countries, and/or 
inaccuracies in reporting. 

Even if the cohort initially tested was only a fraction of the total
infected population, tracking $\bar{M}_{1}(t)$ or $M_{\rm p}^1(t)$ of
this cohort still provides an accurate estimation of the mortality
rate. However, underreporting newly infecteds can confound CFR and
$M_{\rm p}^0(t)$. Current estimates show that only a minority of
SARS-CoV-2 infections are reported (e.g., $f\approx 14\%$ in China
before January 23, 2020)~\cite{LI_SCIENCE}. At early times
(Fig.~\ref{fig:fraction}(a)) most patients, tested or untested, have
not resolved. A reported/tested fraction $f<1$ \textit{would not}
directly affect the CFRs or mortality ratios if the
unreported/untested individuals die and recover in the same proportion
as the tested infecteds (Fig.~\ref{fig:fraction}(b)).  Undertesting
would overestimate the true $M_{\rm p}^0(t)$ and infection fatality
ratio (IFR) if untested (presumably mildly or asymptomatic infected)
individuals are less likely to die than the tested infecteds
(Fig.~\ref{fig:fraction}(c)).  If untested infecteds do not die at
all, the true long-time mortality ${\cal M}_{\rm p}^{0,1}(\infty)
\approx f M_{\rm p}^{0,1}(\infty)$ (see the SI). In the less likely
scenario in which untested individuals do not receive medical care and
hence die at a faster rate than tested patients (see
Fig.~\ref{fig:fraction}(d)), $M_{\rm p}^{0,1}(\infty)$ and CFR based
on the tested fraction would underestimate the true long-time
mortality ${\cal M}_{\rm p}^{0,1}(\infty)$ and IFR, respectively.




Besides underreporting, the delay in transmission after becoming
infected will also affect $M_{\rm p}^0(t)$. Although we have assumed
that transmission occurs only after the incubation period when
symptoms arise, there is evidence of asymptomatic transmission of
coronavirus~\cite{ASYPMTOMATIC,LI_SCIENCE}. Asymptomatic transmission
can be modeled by setting $\beta(\tau) > 0$ even for $\tau < \tau_{\rm
  inc}$. An undelayed transmission in a no-quarantine scenario causes
relatively more new infecteds who have not had the chance to die yet,
leading to a \textit{smaller} mortality ratio $M_{\rm p}^0(t)$.
Within our SIR model, delaying transmission reduces the number of
infected individuals and deaths at any given time but
\textit{increases} the measured mortality ratio $M_{\rm p}^0(t)$.
Without quarantine, the asymptotic values $M_\mathrm{p}^{0}(\infty)$
and ${\rm CFR}(\infty)$ will also change as a result of changing the
transmission latency period, as shown in the SI.  With perfect
quarantining, the asymptote $M_{\rm p}^0(\infty)$ is eventually
determined by a cohort that does not include new infections and is
thus independent of the transmission delay.

In this work, we have explicitly defined a number of interpretable
mathematical metrics that represent the probability of dying from a disease 
%
%
By rigorously defining these metrics, we are able to reveal the
inherent assumptions and factors that affect their estimation. Within
survival probability and SIR-type models, we explicitly illustrate how
physiologically important parameters such as incubation time, death
rate, cure rate, and transmissibility influence the temporal evolution
and asymptotic values of mortality ratios. We also discussed how
statistical factors such as time of testing after infection
($\tau_{1}$) and testing ratio ($f$) affect our estimates.
%
%
In practice, the mortality ratios $M_{1}(t)$ and $M_{\rm p}^{1}(t)$
may provide good estimates of mortality of patients who have tested
positive. In addition to our metrics and mathematical models, we
emphasize the importance of curating individual cohort data. These
data are more directly related to the probability of death $M_{1}(t)$
and are subject to the fewest confounding factors and statistical
uncertainty.

\section*{Data availability}
\vspace{-2mm}

The datasets that we used in this study are stored in the publicly
accessible repositories of
Refs.~\cite{kraemer2020epidemiological,corona1,dong2020interactive}.

\section*{Acknowledgements}
\vspace{-2mm}

LB acknowledges financial support from the SNF Early Postdoc.Mobility
fellowship on ``Multispecies interacting stochastic systems in
biology''. The authors also acknowledge financial support from the
Army Research Office (W911NF-18-1-0345), the NIH (R01HL146552), and
the National Science Foundation (DMS-1814364).

\section*{Competing interests}
\vspace{-2mm}

The authors declare no competing interests.
\section*{Author contributions}
\vspace{-2mm}

LB and TC developed the analyses and wrote the manuscript. LB and MX analyzed data,
performed numerical computations, and edited the manuscript.
\bibliography{refs_TC}

\newpage

\section*{Methods}
\vspace{-2mm}

\subsection*{Numerical scheme}
\vspace{-4mm}

To numerically solve Eqs.~\eqref{SIR1_EQNS} and \eqref{eq:bc_SIR}, we
use a uniform discretization $\tau_{k} = k\Delta{\tau}, k=0,1,\dots,
K$. A backward difference operator $\left[I(\tau_k, t) - I(\tau_{k-1},
  t)\right]/(\Delta{\tau})$ is used to approximate
$\partial_{\tau}{I}(\tau, t)$ and a predictor-corrector Euler scheme
is used to advance time~\cite{RECIPES}. Setting the cut-offs
$I(-\Delta{\tau}, t) \equiv 0$ and $I(K\Delta{\tau}, t) \equiv 0$, the
resulting discretized equations for the full SIR model are
\begin{equation}
\begin{aligned}
S(t+\Delta{t}) = & S(t) - \Delta{t}S(t)\sum_{k=0}^{K}\beta(\tau_k, t) I(\tau_k, t)
\Delta{\tau}, \\
\tilde{I}(\tau_k, t)  = & I(\tau_k, t) - \Delta{t}\f{I(\tau_k, t) 
- I(\tau_{k-1}, t)}{\Delta{\tau}} - \Delta t (c(\tau_k, t) + \mu(\tau_k, t))I(\tau_k, t),\\
I(\tau_k, t+\Delta{t}) = & I(\tau_k, t) -\f{\Delta t}{2}\left[\f{I(\tau_k, t) 
- I(\tau_{k-1}, t)}{\Delta{\tau}} + (c(\tau_k, t) + \mu(\tau_k, t)) 
I(\tau_k, t)\right. \\ 
\: & \quad  + \left. \f{\tilde{I}(\tau_k, t) - \tilde{I}(\tau_{k-1}, t)}{\Delta{\tau}} 
+ (c(\tau_k, t+\Delta t) + \mu(\tau_k, t+\Delta t))\tilde{I}(\tau_k, t)\right] \\
\: & \quad + \delta_{k,0}\frac{\Delta t}{\Delta{\tau}} S(t)
\sum_{j=0}^{K}\beta(\tau_{j},t)I(\tau_{j},t) \Delta{\tau},
%
\label{NUM_ALGO}
\end{aligned}
\end{equation}
where $\tilde{I}$ is the initial predicted guess, and the last term
proportional to $\delta_{k,0}$ encodes the boundary condition
Eq.~\eqref{eq:bc_SIR}. Note that we use $\sum_{k=0}^{K}\beta(\tau_k,
t)I(\tau_k, t) \Delta{\tau}$ to indicate the numerical evaluation of
$\int_0^\infty \mathrm{d}\tau' \beta(\tau',t) I(\tau',t)$. Quadrature
methods such as Simpson's rule and the trapezoidal rule can be used to
approximate the integral more efficiently.

The total deaths, recovereds, and infecteds at time $t$ are found by
\begin{equation}
\begin{aligned}
D_{0}(m\Delta t) & = {1\over 2}\sum_{j=0}^{m}\sum_{k=0}^{K}
c(k\Delta \tau, j\Delta t)\left[I(k\Delta \tau, j\Delta t)+
\tilde{I}(k\Delta \tau, j\Delta t)\right]\Delta \tau 
\Delta{t}, \nonumber \\
R_{0}(t) & = {1\over 2}\sum_{j=0}^{m}\sum_{k=0}^{K}
\mu(k\Delta \tau, j\Delta t)\left[I(j\Delta \tau, j\Delta t)+
\tilde{I}(k\Delta \tau, j\Delta t)\right]\Delta \tau 
\Delta{t}, \nonumber \\
I(m\Delta t) & = \sum_{k=0}^{K} I(k\Delta \tau, m\Delta t)\Delta \tau,
\end{aligned}
\end{equation}
with analogous expressions for $D_{1}(m\Delta t)$ and $R_{1}(m\Delta
t)$. To obtain a stable integration scheme, the time steps $\Delta{t}$
and $\Delta{\tau}$ have to satisfy $\Delta t/(2 \Delta \tau)<1$. In all
of our numerical computations, we thus set $\Delta{t}=0.002,
\Delta{\tau} = 0.02$, and $K=10^{4}$. In the SI, we show additional
plots of the magnitude of $I(\tau, t)$ in the $t-\tau$ plane.

\subsection*{Solutions for $\tau_{1}$-averaged probabilities}
\vspace{-4mm}

Using the method of characteristics, we find the formal solution to
Eq.~\eqref{eq:ind_survival}:

\begin{equation}
P(\tau, t\vert \tau_{1}) = \delta(\tau-t-\tau_{1})e^{-\int_{0}^{t}
(\mu(\tau-t+s,s\vert \tau_{1})+ c(\tau-t+s,s\vert\tau_{1}))\mathrm{d}s},
\label{PATAUT_SOLN}
\end{equation}
which can be used to construct the death and cure probabilities 

\begin{align}
P_{\rm d}(t\vert \tau_{1}) & = \int_{0}^{t}\dd t'\, \mu(\tau_{1}+t', t')
e^{-\int_{0}^{t'}(\mu(\tau_{1}+s,s)+ c(\tau_{1}+s,s))
\mathrm{d}s} \nonumber \\
P_{\rm r}(t\vert \tau_{1}) & = \int_{0}^{t}\dd t'\, c(\tau_{1}+t', t')
e^{-\int_{0}^{t'}(\mu(\tau_{1}+s,s)+ c(\tau_{1}+s,s))\mathrm{d}s}.
\label{PDPC_SI}
\end{align}

If we now invoke  the functional forms of $\mu$ and $c$ given in
Eq.~\eqref{eq:recovery_mortality_rates}, we find explicitly

\begin{equation}
P_{\rm d}(\tau, t\vert \tau_{1})= \left\{ \begin{array}{ll} \displaystyle {\mu_{1} \over
    \mu_{1}+c}\left(1-e^{-(\mu_{1}+c)t}\right) & \tau > t+\tau_{\rm
    inc} \\[13pt] 
0 & \tau_{\rm inc} \geq \tau > \tau_{1} \\[13pt]
 \displaystyle  {\mu_{1}e^{-c(\tau_{\rm inc}-\tau_{1})}\over \mu_{1}+c}\left( 1 -
    e^{-(\mu_{1}+c)(\tau-\tau_{\mathrm{inc}})}\right) & \tau >
  \tau_{\rm inc} \geq \tau_{1}\end{array}\right.
\label{eq:death_prob_analytic}
\end{equation}
and
\begin{equation}
P_{\rm r}(\tau,t\vert \tau_{1}) = \left\{ \begin{array}{ll}
\displaystyle {c \over \mu_{1}+c}\left(1-e^{-(\mu_{1}+c)t}\right) & \tau 
> t+\tau_{\rm inc} \\[13pt]
\displaystyle 1-e^{-ct} & \tau_{\rm inc} \geq \tau > \tau_{1} \\[13pt]
\displaystyle 1-e^{-c(\tau_{\rm inc}-\tau_{1})} + 
{c e^{-c(\tau_{\rm inc}-\tau_{1})}\over \mu_{1}+c}\left(
1 - e^{-(\mu_{1}+c)(\tau-\tau_{\mathrm{inc}})}\right)  & 
\tau > \tau_{\rm inc} \geq \tau_{1}.\end{array}\right.
\label{eq:cure_prob_analytic}
\end{equation}
\begin{figure}[h!]
\centering
\includegraphics[width=2.5in]{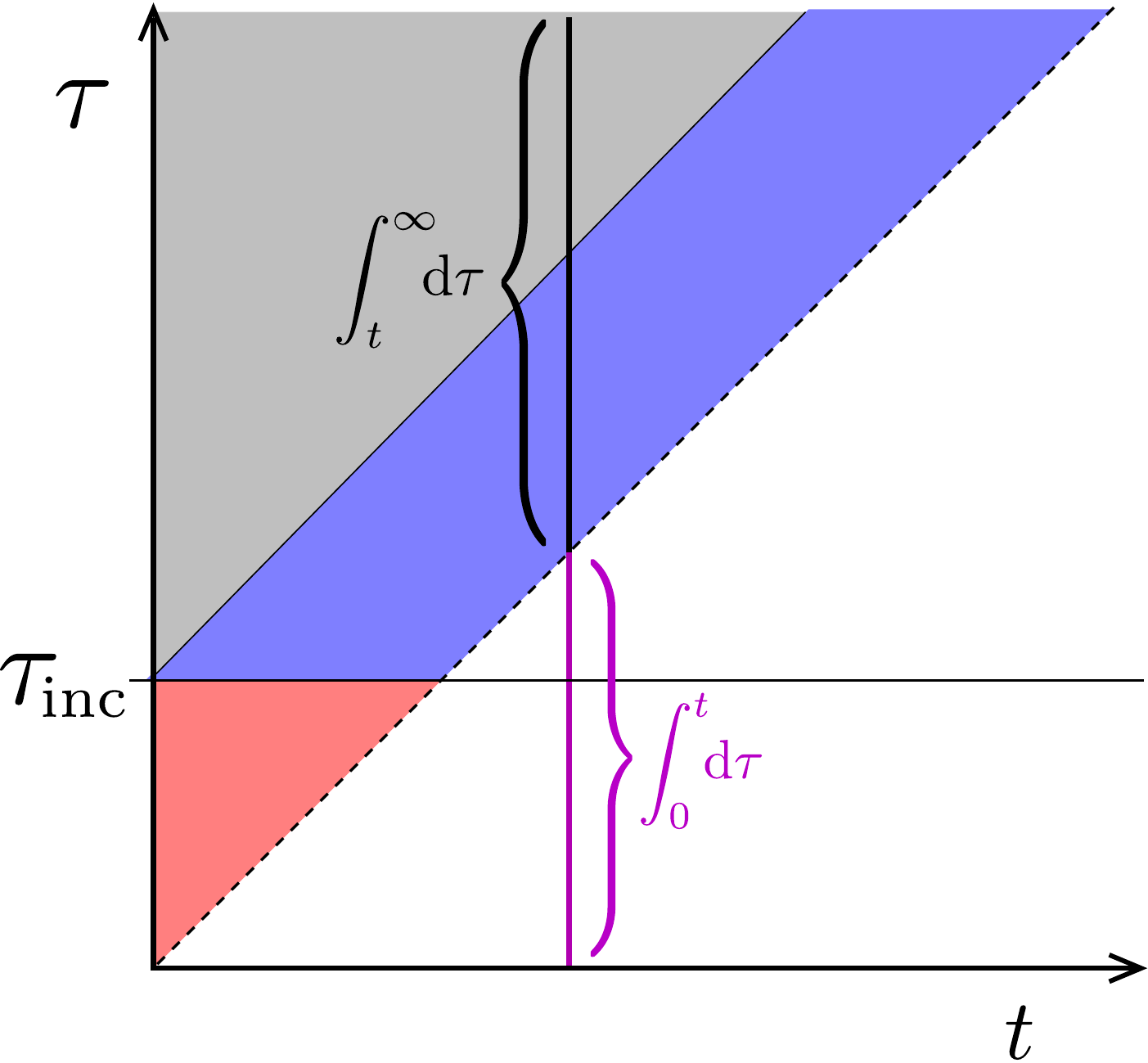}
\caption{\textbf{Phase plot for $P(\tau>t,t)$ and $I(\tau>t,t)$.}  The
  regions delineating different forms for the solution
  (Eq.~\eqref{AVE_P_SOLN}). Here, we have included an incubation time
  $\tau_{\rm inc}$ before which no death occurs.  The solution for
  $\bar{P}(\tau,t)$ or $I(\tau,t)$ in the $\tau < t$ region must be
  self-consistently solved using the boundary condition
  Eq.~\eqref{eq:bc_SIR}. At any fixed time, the integral of
  $I(\tau,t)$ over $t<\tau \leq \infty$ captures only the initial
  population, excludes newly infecteds, and is used to compute
  $D_{1}(t), R_{1}(t)$, and $M_{\rm p}^{1}(t)$. To compute $D_{0}(t),
  R_{0}(t)$, and $M_{\rm p}^{0}(t)$, we integrate across all infecteds
  (including the integral over $t > \tau > \geq 0$ shown in magenta).}
\label{TAU-T}
\end{figure}

Finally, we can also find the $\tau_{1}$-averaged 
probabilities for $\tau \geq t$ by weighting over $\rho(\tau_{1}; n,\gamma)$.
For example, 

\begin{equation}
\bar{P}(\tau, t) = \left\{\begin{array}{ll}
\rho(\tau-t;n,\gamma)e^{-(\mu_{1}+c)t} & \tau \geq t+\tau_{\rm inc} \nonumber \\
\rho(\tau-t;n,\gamma)e^{-ct} & {\color{red}\tau_{\rm inc} \geq \tau > t} \\
\rho(\tau-t;n,\gamma)e^{-ct}e^{-\mu_{1}(\tau-\tau_{\rm inc})}
& {\color{blue}t+\tau_{\rm inc}\geq  \tau >\tau_{\rm inc}} \end{array}\right. .
\label{AVE_P_SOLN}
\end{equation}
These solutions hold for the different regions shown in the phase plot
of Fig.~\ref{TAU-T} and are equivalent to those for
$I(\tau>t,t)$. Corresponding expressions for $\bar{P}_{\rm d}(t)$ and
$\bar{P}_{\rm r}(t)$ can be found and used to construct $M_{\rm
  p}^1(t)$.
%
%
%
%
\begin{figure}[htb]
\centering
\includegraphics[width = 0.85\textwidth]{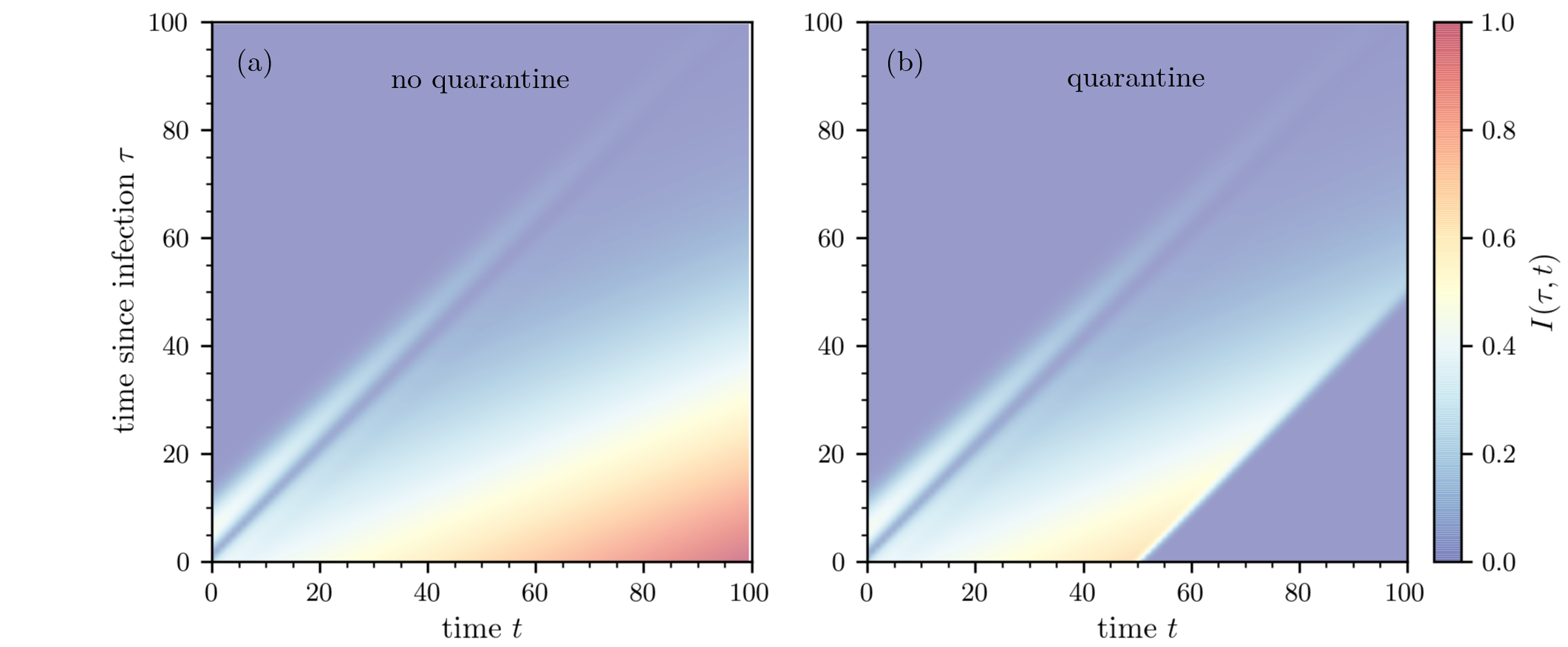}
\includegraphics{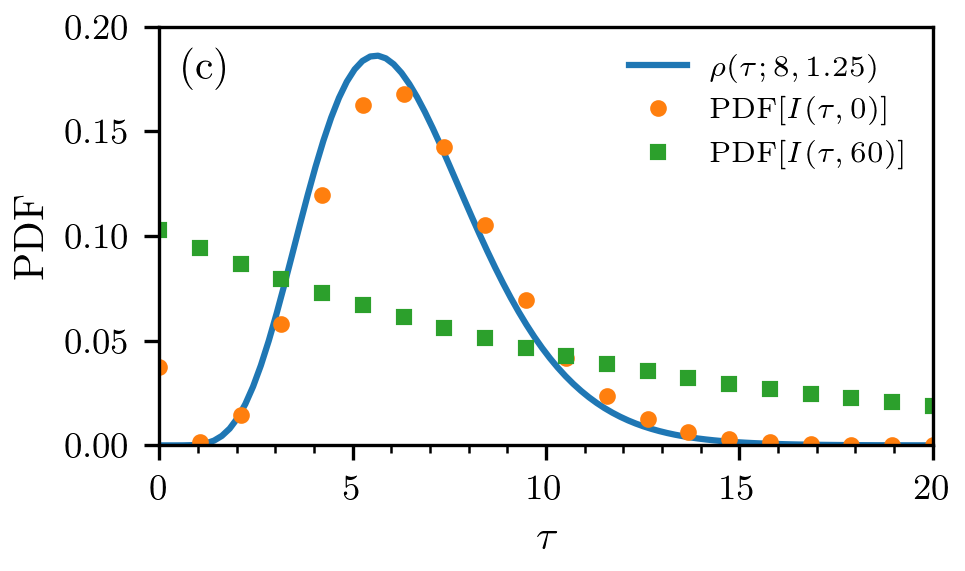}
\includegraphics{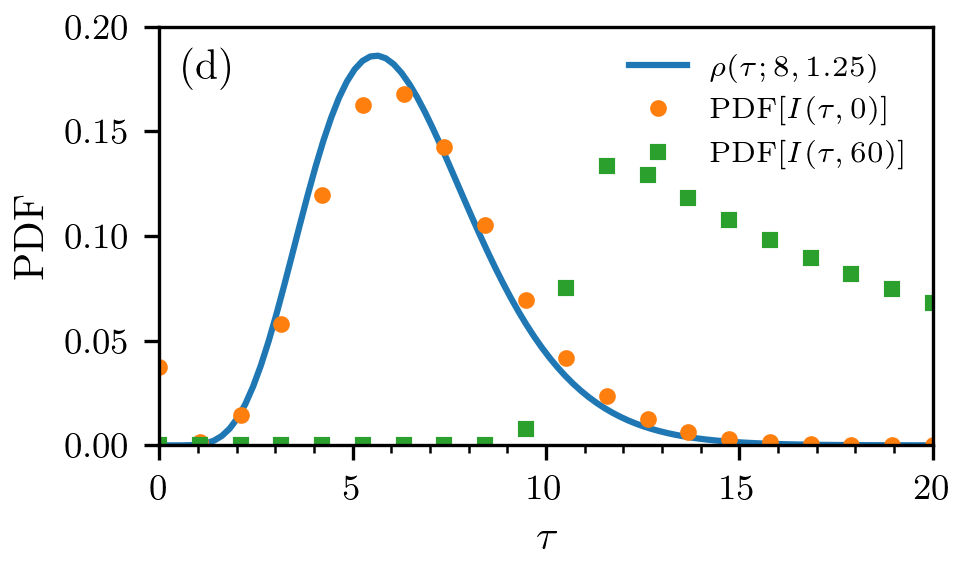}
\caption{\textbf{Density plots of $I(\tau,t)$ in the $t-\tau$ plane.}
  Numerical solution of the equation for $I(\tau,t)$ in
  Eqs.~\eqref{SIR1_EQNS} under the assumption of a fixed susceptible
  size $\beta_{1}S = 0.158$/day.  (a) The density without quarantine
  monotonically grows with time $t$ in the region $\tau < t$ as an
  unlimited number of susceptibles continually produces
  infecteds. (b) With quarantining after $t_{\rm q}=50$ days, we set
  $\beta_{1}S = 0$ for $t > t_{\rm q}$, which shuts off new
  infections. Both plots were generated using the same initial density
  $\rho(\tau_{1})$ defined in Eq.~\eqref{eq:gamma}. In both cases, the
  density $I(\tau > t)$ is identical to $P(\tau> t)$ if the same
  $\rho(\tau_{1})$ is used and is independent of disease transmission,
  susceptible dynamics, etc. (c-d) Probability-density functions
  (PDFs) of the number of infected $I(\tau,t)$ for $t=0,60$ (b)
  without and (c) with quarantine. The blue solid line corresponds to
  the initial distribution $\rho(\tau;n=8,\gamma=1.25)$ (see
  Eq.~\eqref{eq:gamma}).}
\label{DENSITY}
\end{figure}
Fig.~\ref{DENSITY}(a) shows the magnitude of $I(\tau, t)$ in the
$t-\tau$ plane when we set $S(t)=S$ constant (so that the first
equation in Eq.~\eqref{NUM_ALGO} does not apply) such that
$\beta_{1}S \approx 0.158$/day. In this case, the epidemic continues
to grow in time, but the mortality rates $M_{\rm p}^{0,1}(t)$
nonetheless converge as $t\to \infty$. In Fig.~\ref{DENSITY}(b), we
set $\beta_1 S = 0$ for $t > t_{\mathrm{q}}$ to model strict
quarantining after $t_{\mathrm{q}}=50$ days. We observe no new
infections after the onset of strict quarantine measures. In both
cases (quarantine and no quarantine), we use
$\rho(\tau;n=8,\gamma=1.25)$ (see Eq.~\eqref{eq:gamma} in the main
text) to describe the initial distribution of infection times $\tau$.
As time progresses, more of the distribution of $\tau$ moves towards
smaller values until quarantine measures take effect (see
Fig.~\ref{DENSITY}(c) and (d)).

\clearpage

\renewcommand{\theequation}{S\arabic{equation}}
\renewcommand{\thefigure}{S\arabic{figure}} 
\setcounter{figure}{0}   
\setcounter{equation}{0}  
\setcounter{page}{1}

\section*{Supplementary Information}
\vspace{-2mm}

\subsection*{Additional examples of mortality-ratio evolutions}

\vspace{-4mm}
\begin{figure}[htb]
\centering
\includegraphics{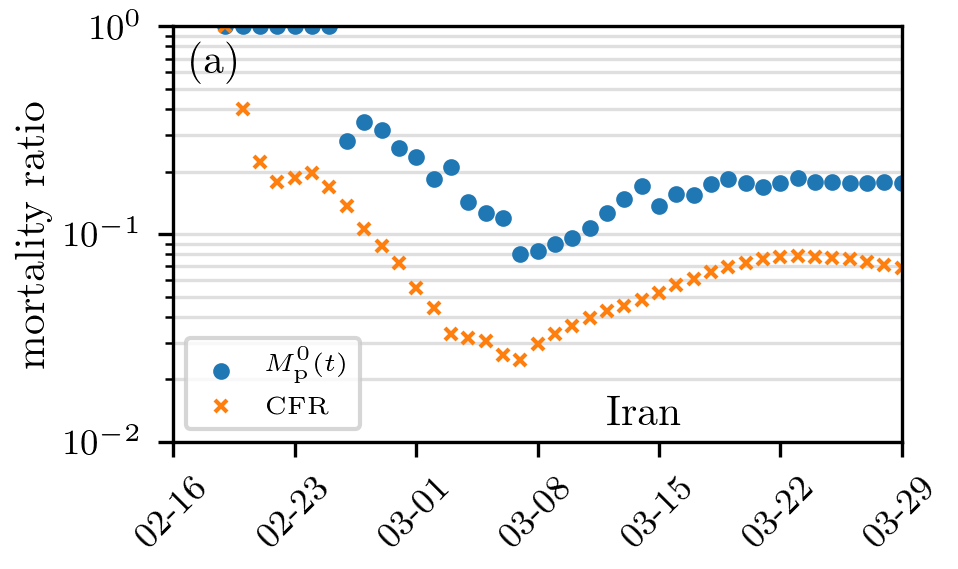}
\includegraphics{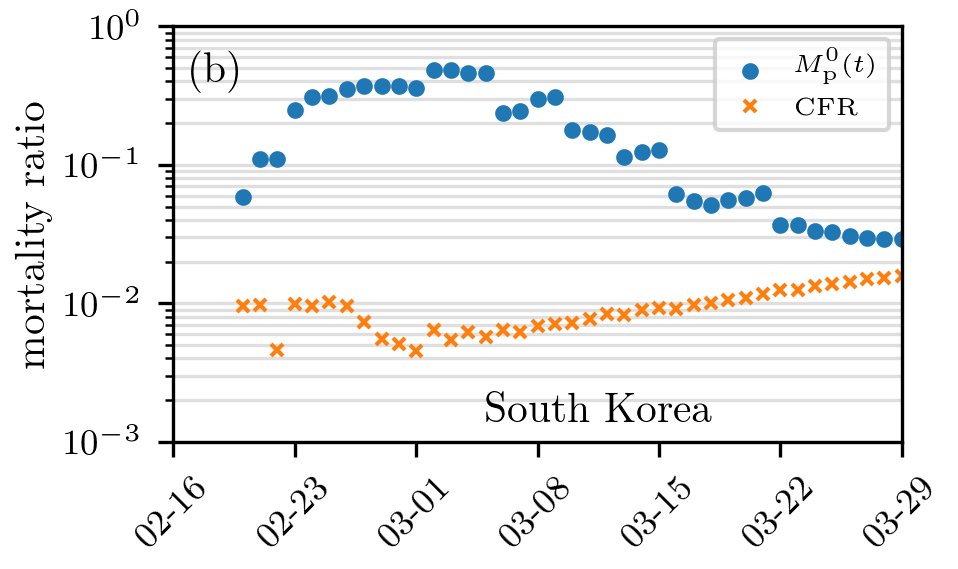}
\includegraphics{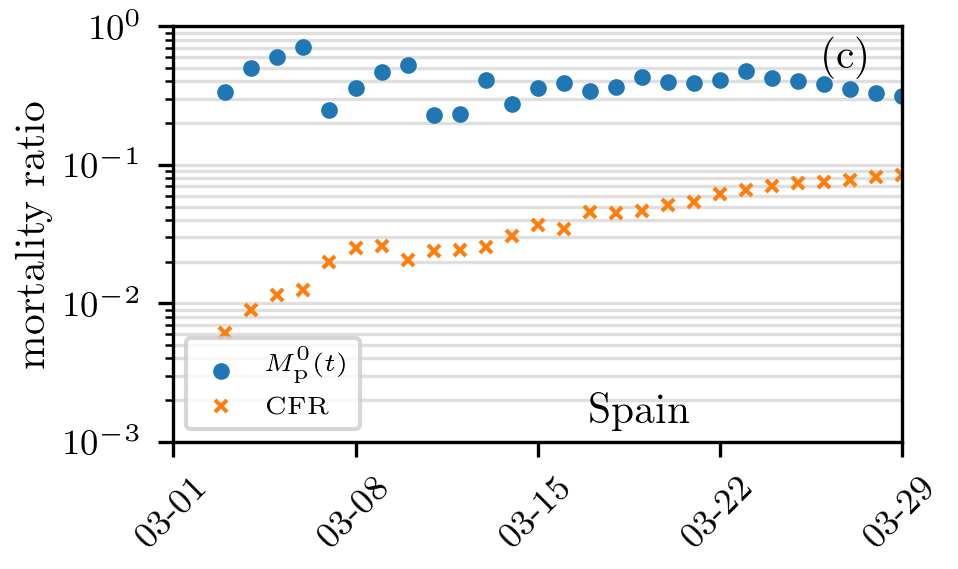}
\includegraphics{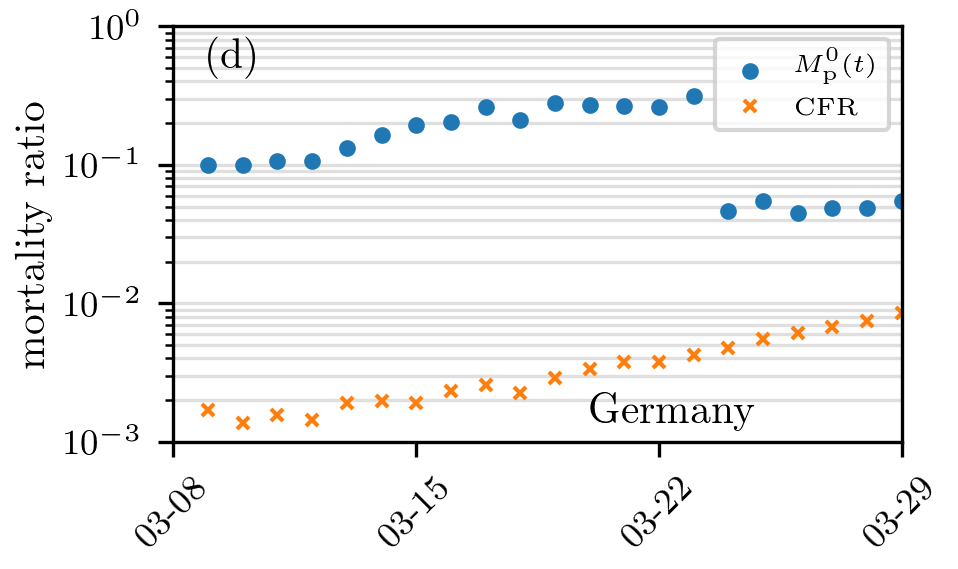}
\includegraphics{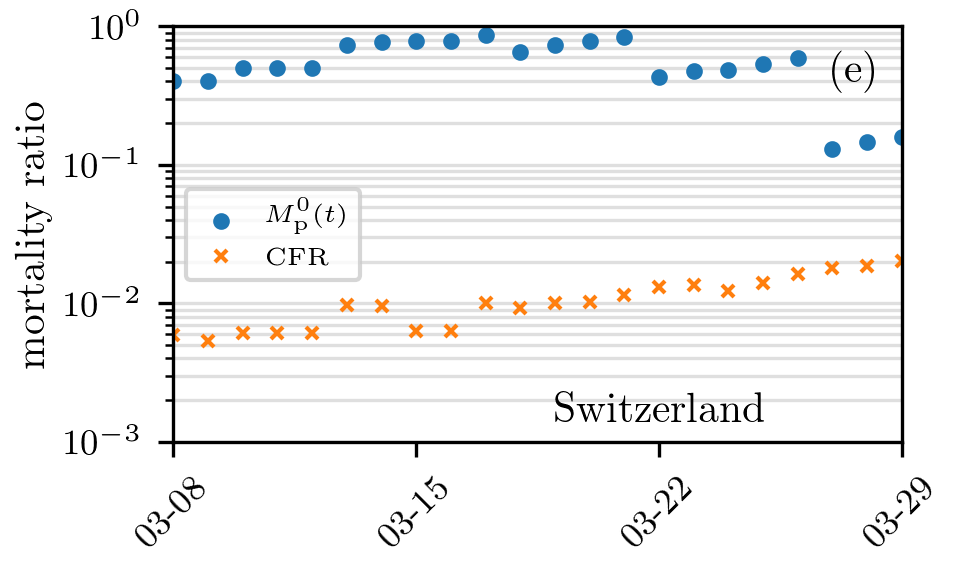}
\includegraphics{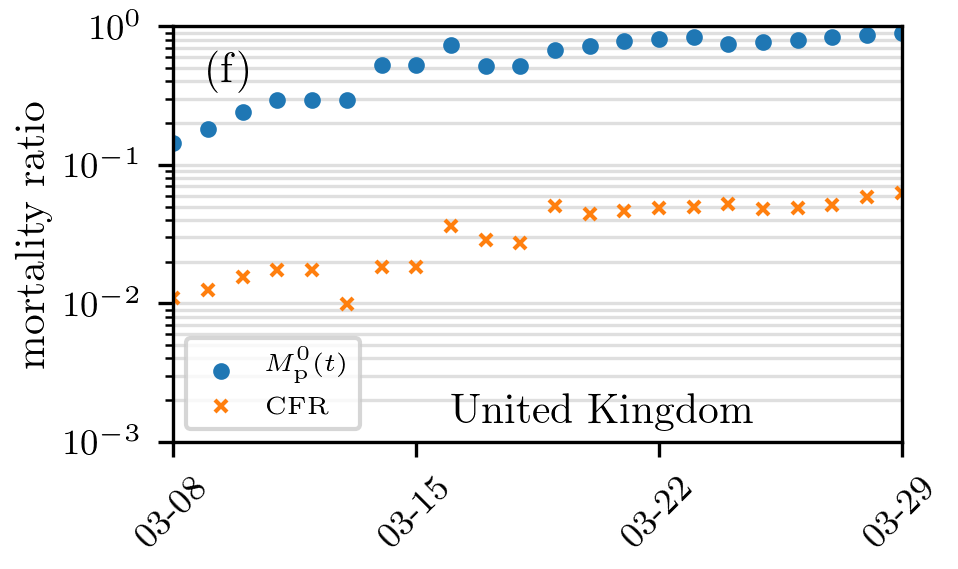}

\caption{\textbf{Mortality ratio estimates}. Estimates of mortality
  ratios (see Eqs.~\eqref{eq:estimate2} and \eqref{eq:Mp0_Mp1} in the
  main text) of SARS-CoV-2 infections in different countries. The case
  fatality rate, $\mathrm{CFR}$, corresponds to the number of deaths to
  date divided by the total number of cases to date.  Another
  population-based mortality ratio is $M_{\mathrm{p}}^0(t)$, the
  number of deaths divided by the sum of deaths and recovereds, up to
  time $t$. The data are derived from Ref.~\cite{dong2020interactive}.}
\label{fig:mortalitydataSI}
\end{figure}
In Fig.~\ref{fig:mortalitydataSI}, we show additional examples of
mortality-ratio estimates for Iran, South Korea, Spain, Germany,
Switzerland, and the United Kingdom. As in
Fig.~\ref{fig:mortalitydata} in the main text, we observe that, by
definition, the population-based mortality ratio $M_\mathrm{p}^0(t)$
is significantly larger than the corresponding CFR in all cases.

\subsection*{Effects of undertesting}
\vspace{-4mm}

Note that $I(\tau,t)$ in the SIR equations determines the dynamics of
the actual infected population. However, (i) typically only a fraction
$f$ of the total number of infecteds might be tested and confirmed
positive and (ii) the testing of newly infecteds may also be delayed
by a distribution $\rho(\tau; n,\gamma)$.

If positive tests represent only a fraction $f$ of the total infected
population, and the confirmation of newly infecteds occurs
immediately, the known infected density is given by $I^{*}(\tau,t) =
fI(\tau,t)$ where $I(\tau,t)$ is the true total infected
population. If testing of newly infecteds occurs after a distribution
$\rho(\tau;n,\gamma)$ of infection times, $I^{*}(\tau,t) = f
\int_{0}^{\tau}I(t-\tau+\tau_{1},
t)\rho(\tau_{1};n,\gamma)\dd\tau_{1}$.

In our development of $M_{\rm p}^{0,1}(t)$
and ${\rm CFR}_{\rm d}(t,\tau_{\rm res})$ in the manuscript, we
assumed the entire infected population was tested and confirmed. Thus,
$M_{\rm p}^{0,1}(t)$ and ${\rm CFR}_{\rm d}(t,\tau_{\rm res})$ were
computed using $f=1$ and more accurately represent the mortality
ratios of the population \textit{conditioned} on being tested
positive.

\begin{figure}[htb]
\centering
\includegraphics[width=6.9in]{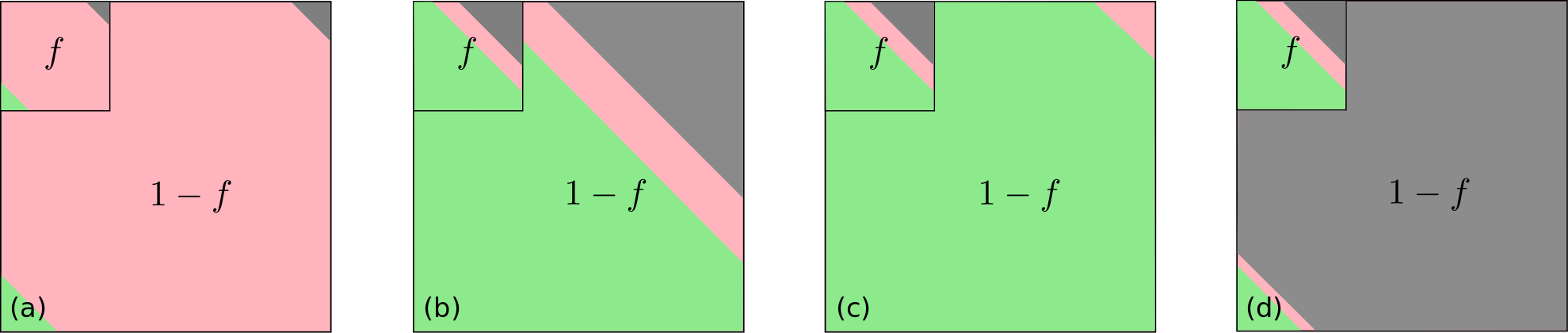}
\caption{\textbf{Fractional testing.} An example of fractional testing
  in which a fixed fraction $f$ of the real total infected population
  is assumed to be tested. The remaining $1-f$ proportion of infecteds
  are untested. Equivalently, if the total tested fraction has unit
  population, then the total population of the untested pool is
  $1/f-1$. (a) At short times after an outbreak, most of the infected
  patients, tested and untested, have not yet resolved (red). Only a
  small number have died (gray) or have recovered (green). (b) At
  later times, if the untested population dies at the same rate as the
  tested population, $M_{\rm p}(t)$ and CFR remain accurate estimates
  for the entire infected population. (c) If the untested population
  is, say, asymptomatic and rarely dies, the true mortality ${\cal
    M}_{\rm p}^{0,1}(t) \approx f M_{\rm p}^{0,1}(t)$ is overestimated
  by the tested mortality $M_{\rm p}^{0,1}(t)$. (d) Finally, in a
  scenario in which untested infecteds die at a higher rate than tested
  ones, $M_{\rm p}^{0,1}(t)$ and CFR based on the tested fraction
  \textit{underestimate} the true mortalities.}
\label{fig:fraction}
\end{figure}

To estimate the mortality ratio of the population conditioned simply
on being infected, we have to estimate the larger number of recovereds
that went untested.  For the most likely scenario in which untested
infecteds have negligible death rate, as shown in
Fig.~\ref{fig:fraction}(c), we can employ the SIR model without death
for the untested pool of infecteds,

\begin{align}
{\dd S(t)\over \dd t}  & =  
-S(t)\int_{0}^{\infty}\!\dd\tau'\, \beta(\tau',t)(I^{*}(\tau',t)+ 
I^{\rm u}(\tau',t)),\nonumber \\
{\partial I^{*}(\tau, t)\over \partial t}+{\partial I^{*}(\tau,t)\over
  \partial \tau} & =
-(\mu(\tau,t)+c(\tau,t))I^{*}(\tau,t), \nonumber \\
{\partial I^{\rm u}(\tau, t)\over \partial t}+{\partial I^{\rm u}(\tau,t)\over
  \partial \tau} & =
- c(\tau,t)I^{\rm u}(\tau,t), \nonumber \\
{\dd R(t)\over \dd t} & = \int_{0}^{\infty}\!\dd \tau c(\tau,t)
(I^{*}(\tau,t)+I^{\rm u}(\tau,t)),
%
\label{SIR1_TU_EQNS}
\end{align}
where $I^{*}(\tau,t)+I^{\rm u}(\tau,t)=I_{\rm T}(\tau, t)$, the total
density of infecteds, and the production of tested and untested
infecteds follow the boundary conditions

\begin{align}
I^{*}(0,t) & = f S(t) \int_{0}^{\infty}\!\dd\tau\, \beta(\tau,t)I_{\rm T}(\tau,t) \nonumber \\
I^{\rm u}(0,t) & = (1-f) S(t) \int_{0}^{\infty}\!\dd\tau\, \beta(\tau,t)I_{\rm T}(\tau,t).
\end{align}
Here, we have assumed that testing occurs only for the infected who
can die. In this scenario, the IFR, or the true CFR of all infecteds,
is then ${\rm IFR}(t) = D_{0}^{*}(t)/N_{\rm T}(t)$, where in analogy
to Eqs.~\eqref{eq:Nmu0_Nc0} and \eqref{eq:NT0},

\begin{equation}
D_0^{*}(t) = \int_{0}^{t}\!\!\dd t' \int_{0}^{\infty}\!\!\dd \tau\,  
\mu(\tau, t')I^{*}(\tau,t'),\quad  R_0(t) =
\int_{0}^{t}\!\!\dd t'\int_{0}^{\infty}\!\!\dd \tau\,
c(\tau, t')I_{\rm T}(\tau,t'),
\label{eq:Nmu0_Nc0_F}
\end{equation}
and 
\begin{equation}
N_{\rm T}(t)=D_0^{*}(t) + R_{0}(t) + \int_{0}^{\infty}\!\!\dd \tau\,
I_{\rm T}(\tau, t).
\label{eq:NT0_F}
\end{equation}
The true mortality ratio is also straightforwardly defined by, for
example,

\begin{equation}
{\cal M}_{\rm p}^0(t) = {D^{*}_{0} \over D_{0}^{*}(t) + R_{0}^{*}(t) + R_{0}^{\rm u}(t)},
\end{equation}
where 

\begin{equation}
R_{0}^{*}(t) =  \int_{0}^{\infty}\dd\tau \int_{0}^{t}\dd t'\, c(\tau, t')I^{*}(\tau, t')
\,\,\,\mbox{and}\,\,\, 
R_{0}^{\rm u}(t) = \int_{0}^{\infty}\dd\tau \int_{0}^{t}\dd t'\, c(\tau, t')
I^{\rm u}(\tau, t'),
%
\label{DR_TU}
\end{equation}
with analogous expressions for $D_{1}^{*}(t)$, $R^{*}_{1}(t)$, and 
$R^{\rm u}_{1}(t)$. At long times, after resolution of all infecteds, 
the untested recovered population is

\begin{equation}
R^{\rm u}_{0,1}(\infty) = \left({1\over f} -1\right)(D_{0,1}^{*}(\infty) + R_{0,1}^{*}(\infty)),
\end{equation}
which yields the asymptotic true ratio ${\cal M}_{\rm p}^{0,1}(\infty)
= fM_{\rm p}^{0,1}(\infty)$ as described in the Discussion and
Summary. In this simple rescaling to account for untested populations,
we have assumed that all deaths come from the tested pool and that the
recovery rate $c$ is the same in the tested and untested pools.


\subsection*{Influence of different transmission rates}
\vspace{-4mm}

\begin{figure}[htb]
\centering
\includegraphics{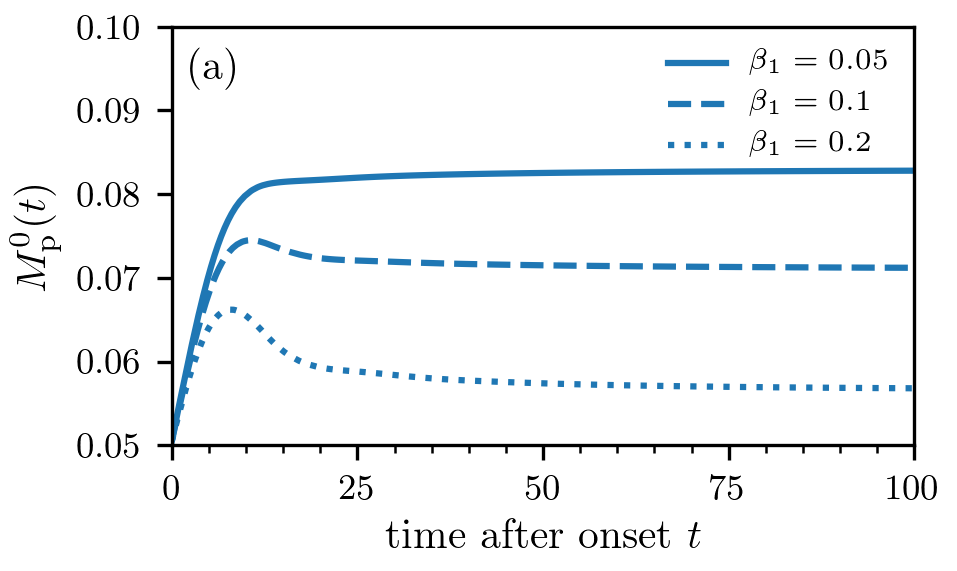}
\includegraphics{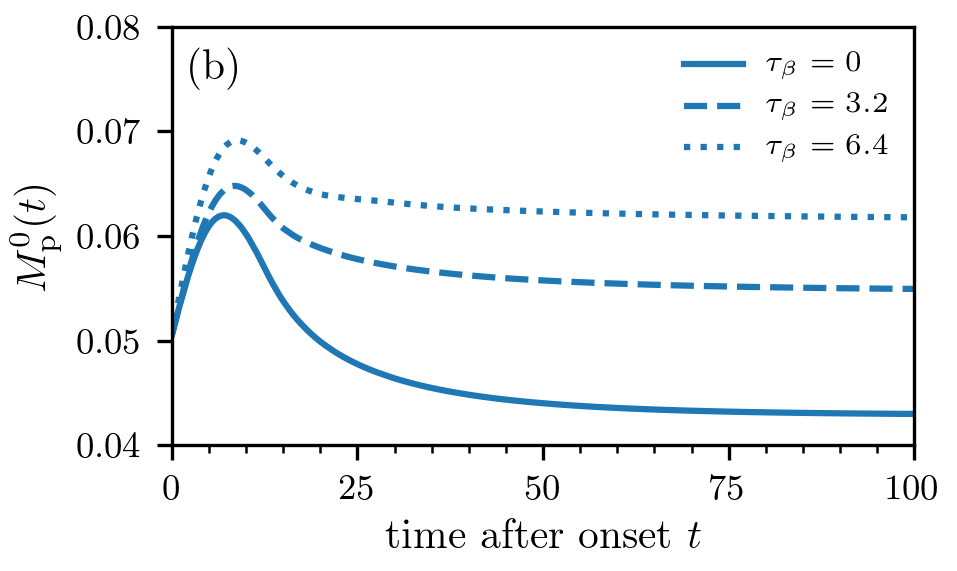}
\caption{\textbf{Population-level mortality for different infection
    rates.} (a) The population-level mortality ratio
  $M_{\mathrm{p}}^0(t)$ for different values of $\beta_1$ and an
  incubation time of $\tau_{\rm inc} = 6.4$ days. In the initial
  exponential growth phase of the epidemic (\textit{i.e.},
  $S(t)\approx S_0$), larger infection rates $\beta_1$ lead to smaller
  values of $M_{\mathrm{p}}^0(t)$. (b) We observe a similar effect for
  non-delayed transmissions (\textit{i.e.}, $\tau_{\beta}\approx
  0$). As long as $S(t)\approx S_0$, smaller transmission delays
  $\tau_{\beta}$ lead to larger relative numbers of new infections and
  smaller $M_{\mathrm{p}}^0(t)$.}
\label{fig:delay}
\end{figure}

In Fig.~\ref{fig:sir} of the main text, we observe that the
population-level mortality ratio $M_{\mathrm{p}}^0(t)$ approaches a
plateau during the initial exponential growth phase of an epidemic
(\textit{i.e.}, for $S(t)\approx S_0$). If the number of new
infections decreases (\textit{e.g.}, due to quarantine measures),
$M_{\mathrm{p}}^0(t)$ starts growing until it reaches its asymptotic
value $M_{\mathrm{p}}^0(\infty)$. Interestingly, the pre-asymptotic
values of $M_{\mathrm{p}}^0(t)$ are smaller for larger infection rates
$\beta_1$ (see Fig.~\ref{fig:delay}(a)). This counter-intuitive effect
arises because larger values of $\beta_1$ generate relatively larger
numbers of new infected which have a lower chance of dying before
$\tau_{\rm inc}$ (see Eq.~\eqref{eq:recovery_mortality_rates} in the
main text).
A similar effect occurs for non-delayed transmission (\textit{i.e.},
$\tau_{\beta} \approx 0$). As the transmission delay decreases, more
secondary cases will result from one infection, leading to 
smaller values of $M_{\mathrm{p}}^0(t)$ in the initial exponential
growth phase of an epidemic (see Fig.~\ref{fig:delay}(b)).

%
\end{document}